# Interplay of anionic quasi-atom and interstitial point defects in electrides: abnormal interstice occupation and colossal charge state of point defects in dense FCC-lithium

L. Zhang,[†] Q. Wu,[†] S. Li,[†] Y. Sun,[†] X. Yan,[†,‡] Y. Chen,[+] and Hua Y. Geng[*,†,§]

[†]*National Key Laboratory of Shock Wave and Detonation Physics, Institute of Fluid Physics, CAEP, P.O. Box 919-102, Mianyang, Sichuan 621900, P.R. China*

[‡]*Jiangxi University of Science and Technology, Ganzhou, Jiangxi 341000, P.R. China*

[+]*Fracture and Reliability Research Institute, School of Engineering, Tohoku University 6-6-01 Aramakiaoba, Aoba-ku, Sendai, 980-8579, Japan*

[§]*Center for Applied Physics and Technology, HEDPS, and College of Engineering, Peking University, Beijing 100871, P.R. China*

**Abstract**: Electrides are an emerging class of materials with highly-localized electrons in the interstices of a crystal that behave as anions. The presence of these unusual interstitial quasi-atom (ISQ) electrons leads to interesting physical and chemical properties, and wide potential applications for this new class of materials. Crystal defects often have a crucial influence on the properties of materials. Introducing impurities has been proved to be an effective approach to improve the properties of a material and to expand its applications. However, the interactions between the anionic ISQs and the crystal defects in electrides are as yet unknown. Here, dense FCC-Li was employed as an archetype to explore the interplay between anionic ISQs and interstitial impurity atoms in this electride. This work reveals a strong coupling among the interstitial impurity atoms, the ISQs, and the matrix Li atoms near to the defects. This complex interplay and interaction mainly manifest as the unexpected tetrahedral interstitial occupation of impurity atoms and the enhancement of electron localization in the interstices. Moreover, the Be impurity occupying the octahedral interstice shows the highest negative charge state ($Be^{8-}$) discovered thus far. These results demonstrate the rich chemistry and physics of this emerging material, and provide a new basis for enriching their variants for a wide range of applications.

**Keywords**: Electride; Interstitial quasi-atom (ISQ); Electron localization; Point defects; Dense lithium

---

[*] To whom correspondence should be addressed. E-mail: s102genghy@caep.cn

# 1 Introduction

Inorganic electrides form a class of emerging materials[1] that has attracted considerable attention since the synthesis of the first room-temperature stable inorganic electride of $[Ca_{24}Al_{28}O_{64}]^{4+}(e^-)_4$ in 2003[2]. Recent first-principles calculations predict that there may be approximately 200 potential inorganic electride materials[3-8]. In particular, high pressure has been proved to be an efficient way to enhance the electron localization in the lattice interstice and to synthesize novel electrides[9-12]. Another aspect that should be noted is that these potential inorganic electrides mainly contain elemental metals and alloys. In other words, electron localization is a very important phenomenon, and cannot be ignored, especially in metals and alloys under compression.

The unique feature of electrides is the presences of unconventional excess electrons, as shown in Fig. 1, which are confined and localized in the interstitial lattice sites and behave as anions[3,13-16]. These confined and localized electrons are usually referred to as interstitial quasi-atoms (ISQ)[3] to emphasize their anionic characteristics. The availability of loosely bonded anionic ISQs in electrides provides interesting physical and chemical properties. For example, their ultra-low work functions[17,18] and high electron mobility[2,19,20] can be applied in catalysts[21-23], novel electrode materials[24,25] and superconductor[26,27]. It has also been shown that these localized anionic ISQs can create an energy gap in metals, leading to counterintuitive metal-nonmetal transitions, such as in Li at ~80 GPa[28,29] and in Na at ~200 GPa[30,31]. This unusual phenomenon is driven by the transfer of $s$ electrons to the $p$ (or $d$) orbits, and the subsequent electron localization and $s$-$p$ (or $p$-$d$) hybridization. The complex structural phase transition in dense Li (from $oC$88 to $oC$40) is attributed to the highly-localized anionic electrons driven by compression, also[29].

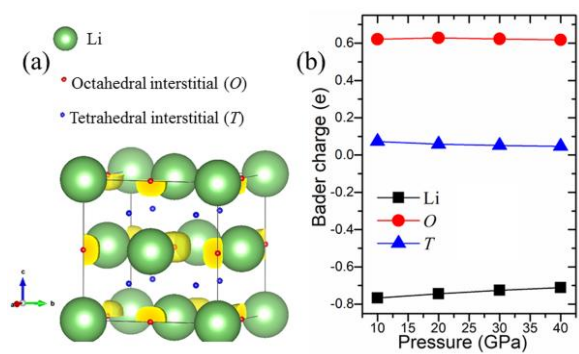

**Figure 1. (Color online) (a) Electron localization function (ELF) of FCC-Li at 10 GPa**

**(isosurface=0.75). The big green, small red and small blue spheres represent Li atom, octahedral interstitial (*O*) and tetrahedral interstitial (*T*) sites, respectively. (b) The variation of Bader charge of Li atoms and ISQ at *O* and *T* sites with pressure, respectively.**

Crystal defects often have a crucial influence on the properties of materials[32-35]. For example, extensive theoretical and experimental works have shown that doping impurities[36-38] and introducing atomic vacancies[39,40] are effective ways to manipulate the electronic state to improve the activity of catalysts[41]. This offers the possibility of tuning or manipulating the localization and delocalization characteristics of electrons in electrides by introducing defects, in order to engineer and enrich variants of these materials for different applications. However, the interaction and interplay between the anionic ISQs and crystal defects in electrides has not been researched previously, due to the short research history of this area. On the other hand, as is well known, electronegativity describes the ability of an atom to attract shared electrons to itself; the existence of interstitial impurities with different values of electronegativity must have a non-negligible influence on the anionic ISQs, and this may induce interesting physics and chemistry in this class of materials. All of these motivate our interest in pursuing solutions to this problem.

It is evident that dense lithium in the face-centered cubic (FCC) phase displays well-defined zero-dimensional (0-D) electride characteristics, with excess electrons localized at *O* sites, as shown in Fig. 1. The variation in the charge on lattice Li atoms and all ISQs with pressure is small, and the stability pressure range of this FCC phase (~10-40 GPa) is easy to achieve in experiments. In this work, dense FCC-Li is therefore chosen in order to facilitate a quantitative analysis of the interplay between the interstitial point defects (the simplest crystal defects) and matrix Li atoms, and ISQs. Three elements with different electronegativity[42-44], He, F, and Be, are selected as doping atoms. In addition, the self-interstitial defects are explored.

## 2 Methodology and computational details

### 2.1 Computational details

First-principles calculations are carried out using a density functional theory (DFT),[45,46] implemented in the Vienna Ab-initio Simulation Package (VASP),[47-49] with the PAW pseudo-potential method and a plane-wave basis set. The generalized-gradient approximation (GGA) of Perdew-Burke-Ernzerhof (PBE) is used for the electronic exchange-correlation functional[50]. The $1s^22s^1$, $1s^2$, $2s^2$, and $2s^22p^5$ electrons are included in the valence space for the Li, He, Be, and F atoms, respectively. An energy cutoff of 900 eV is used for expansion of the plane-wave basis set. A 3×3×3 supercell of the conventional cubic cell of the FCC phase is used as the matrix model (containing 108 Li atoms, 108 $O$ sites and 216 $T$ sites). A point defect is introduced by adding one impurity atom to the $O/T$ interstitial site. Monkhorst-Pack (MP) meshes of 5×5×5 are employed to sample the $k$-points in the Brillouin zone (BZ). The total energy is guaranteed to converge to within 1 meV/atom. All structures investigated here are fully relaxed at the given pressure until the Hellmann-Feynman forces acting on all atoms are less than 0.001 eV/Å, and the total stress tensor is converged to the hydrostatic state within 0.01 GPa. The convergence of the total energy with respect to the $k$-point meshes, supercell size, and the energy cutoff are given in the Supporting Information (SI) (see Figs. S1-S3). For 1×1×1 cubic FCC-Li cells with different densities, KSPACING was set to 0.1 Å$^{-1}$ to ensure the reliability of the results. All Electron localization function (ELF) and charge density are visualized with the VESTA 3 program[51].

A defect was introduced by adding one impurity atom to the interstitial site. What we focused on in this study is that how the point defect altered the local electronic localization characteristics of the matrix. Since the defective system is in a neutral and metallic state within the pressure range studied here, we do not need to consider the global charge compensation as in defective semiconductor and insulator systems[52].

**2.2 Charge state analysis method**

Considering the unique electron localization in electrides, a Bader charge analysis method[53-56], based purely on the electronic charge density, was applied to characterize the charge state of the matrix Li atom, the (self-) interstitial atom, and the ISQs of the systems studied in this work. To reduce calculation errors, a fine fast Fourier transform (FFT) mesh (224×224×224) was adopted to divide the charge density. A statistical analysis method was also used to extract the average value and the standard deviation in the calculated values arising from the introduction of the defect.

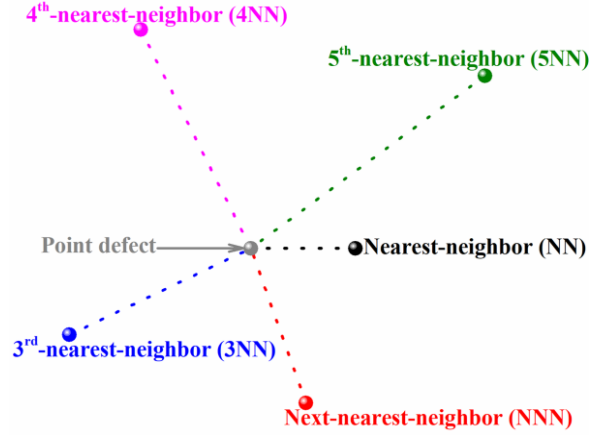

**Figure 2. (Color online) Schematic of the distance between point defects and the matrix Li atoms/ISQs.**

As shown in Fig. 2 (as well as Figs. S4-S8 in SI), for each defective system, we group the nearest- (NN), next-nearest- (NNN), third-nearest- (3NN), and fourth-nearest-neighbor (4NN) matrices of Li atoms for the point defect, and then calculate their Bader charge and statistical averages. The results are denoted as Li(NN), Li(NNN), Li(3NN), and Li(4NN), respectively. The Bader charges on all other matrix Li atoms far from the point defect are also grouped and averaged, and the result is denoted as Li(M). The Bader charges for the ISQs are analyzed in a similar way. Taking a system in which an interstitial point defect He occupies $O$ site (denoted as $I_{He}^{O}$) as an example, these $T$ sites are grouped and labeled as ISQ$^T$(NN), ISQ$^T$(NNN), ISQ$^T$(3NN), and so on, according to the distance between the interstitial He and $T$ sites. The standard deviation of each grouped atoms or ISQs are also calculated. In this work, all atoms and ISQs both within and beyond the influenced area of the defect are carefully analyzed (see Figs. S4-S9 and Tables S1- S6 in SI). The data of the averaged Bader charge and its standard deviation of all studied defects are summarized in the SI (see Tables S8- S15).

For brevity, in following sections, the systems of interstitial point defects (He, Be, F, and Li) occupying the $O/T$ sites in the FCC-Li supercell are denoted as $I_{He}^{O}/I_{He}^{T}$, $I_{Be}^{O}/I_{Be}^{T}$, $I_{F}^{O}/I_{F}^{T}$, and $I_{Li}^{O}/I_{Li}^{T}$, respectively. Only the Bader charge variation of the NN- and NNN-atoms/ISQs of the defect are specifically discussed and compared with the perfect FCC-Li. The detailed pressure dependence of Bader charges of the matrix atoms, impurity atom, and ISQs are listed and shown in the SI (see Figs. S11- S16).

## 3 Results and Discussion

### 3.1 Impact to interstice occupation by localized interstitial electrons

Defect formation enthalpy ($\Delta H$) is an important criterion of thermodynamic stability at finite pressure, and is used in this work to evaluate the relative stabilities of point defects occupying different interstitial sites. Since the defective system Li$mX$ (interstitial atom $X$=He, Be, F, Li) is in a neutral and metallic state within the studied pressure range, $\Delta H$ for defective Li$mX$ can be calculated as follows[57]:

$$\Delta H(\text{Li}mX) = H(\text{Li}mX) - mH(\text{Li}) - H(X) \quad (1)$$

where $H(\text{Li}mX)$ is the enthalpy of the defective system of Li$mX$, and $H(\text{Li})$ and $H(X)$ are the enthalpy per atom of the most favored structure of Li and $X$ at the given pressure, respectively. Based on their respective phase diagrams at the relevant pressure range, the FCC, HCP, HCP, and Cmca structures are used for Li[58], He[59], Be[60], and F[61], respectively.

Considering the source of the impurities (He, F, and Be) may not come from elementary solid, $\Delta H$ is also calculated by the definition

$$\Delta H(\text{Li}mX) = H(\text{Li}mX) - (m-n)H(\text{Li}) - H(\text{Li}_nX) \quad (2)$$

where $H(\text{Li}mX)$ is the enthalpy of the defective system of Li$mX$, $H(\text{Li})$ is the enthalpy per atom of the FCC-Li, and $H(\text{Li}_nX)$ is the enthalpy of Li$_nX$, respectively. $P6/mmm$-LiBe$_2$ (can stable when pressure above 20 GPa)[62] and rocksalt LiF (are stable at the pressure range we studied)[63] are selected as the impurities source of Be and F, respectively (No stable Li$_n$He compounds were reported as far as we know). From Table I we can see that the source of impurities has an obvious influence on the $\Delta H$ of F when compared with that of Be, but it not alters the relatively stability of these defects.

**TABLE I.** Defect formation enthalpy ($\Delta H$, in eV) calculated using different methods (i.e. formula (1) and (2), respectively).

| Pressure (GPa) | $I_F^T$ | $I_F^O$ | $I_{Be}^T$ | $I_{Be}^O$ | Method |
|---|---|---|---|---|---|
| 10 | -5.14 | -5.96 | 1.66 | 1.71 | (1) |
|    | 1.60  | 0.76  | -    | -    | (2) |
| 20 | -5.14 | -6.34 | 1.49 | 1.70 | (1) |
|    | 2.09  | 0.88  | 1.55 | 1.76 | (2) |
| 30 | -5.14 | -6.60 | 1.34 | 1.67 | (2) |
| | 2.46 | 0.99 | 1.45 | 1.78 | (2) |
| 40 | -5.16 | -6.83 | 1.17 | 1.61 | (1) |
| | 2.76 | 1.09 | 1.34 | 1.78 | (2) |

Considering $\Delta H$ is only be used to evaluate the relative stabilities of these point defects in this study, thus $\Delta H$ of these defects are calculated and compared via method (1) in the following to ensure the consistency.

**TABLE II. Calculated formation energy of self-interstitial point defect in typical FCC metals at ambient conditions.**

| FCC metals | Al | Cu | Ag | Au | Ref. |
|---|---|---|---|---|---|
| $E_O^f$ (eV) | 2.44 | 3.28 | 3.05 | 2.73 | 64 |
| | | 3.40 | 3.04 | 2.91 | 65 |
| $E_T^f$ (eV) | 4.77 | 6.64 | 6.30 | 5.54 | 64 |
| | | 4.66 | 4.16 | 4.05 | 65 |

In a perfect FCC structure, $O$ sites have a larger space than the $T$ sties, as shown in Fig. 1a. Therefore, as listed in Table II, the $O$ site is more favored by interstitial atoms in most conventional FCC metals. However, as shown in Fig. 3, interstitial Be and self-interstitial Li atoms are more inclined to occupy the $T$ sites in dense FCC-Li at 10-40 GPa, in sharp contrast to the common wisdom. It is interesting to note that, there is a transition for the interstitial He atom in FCC-Li driven by pressure: at low pressure (< 20 GPa), the $T$ site is favored, whereas it transitions to the $O$ site at higher pressure. This switch in the preferred defect sites of the He impurity could lead to a pseudo-transition in the thermodynamics of the defective FCC-Li phase.

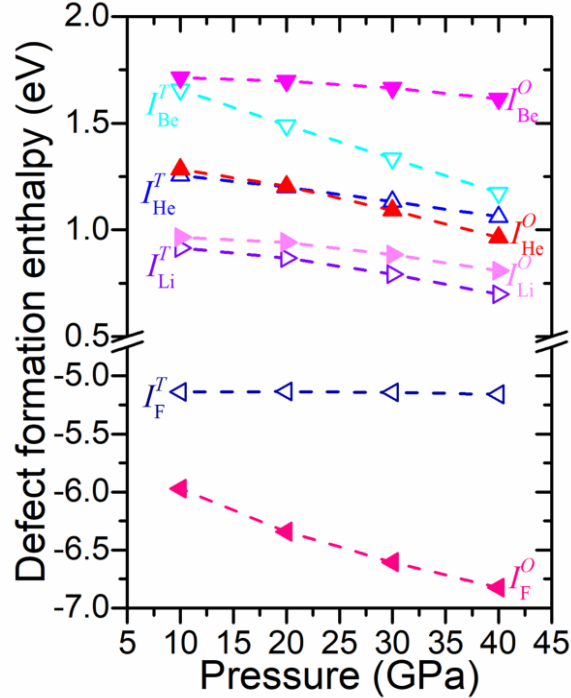

**Figure 3.** (Color online) Calculated defect formation enthalpy in Li$m$X (X=He, F, Be, and self-interstitial Li) as a function of pressures. $I_X^Y$ (X=He, Be, F, Li; Y=O, T) represent interstitial point defect (He, Be, F, and Li) occupying the *O*/*T* interstitial sites, respectively.

This unexpected behavior is a unique behavior of the electride, and can be understood only when the interstitial-localized electron is taken into account. As shown in Fig. 1, the valance electrons in FCC-Li will be localized in the *O* sites under pressure. This anionic ISQ has a strong repulsive interaction with atoms with low electronegativity, whereas it has a strong attractive interaction (or bonding) with atoms with high electronegativity. This explains the low formation energy of F at the *O* site when compared with that of Be. The same mechanism also explains the change in the preferred site of the He impurity in FCC-Li at ~20 GPa.

Due to this mechanism, the preferred interstitial occupation in FCC-Li will change with the redistribution of the localized electrons. We validate this hypothesis by checking the calculated enthalpy of defective FCC-Li, using a 1×1×1 cubic cell. As shown in Fig. 4, the ELF undergoes an obvious change with variation in the density of perfect FCC-Li. In particular, the electrons have a nearly homogeneous distribution in the crystal interstices when $\rho/\rho_0$=0.75. According to the definition of the ELF[66], we can say that electride FCC-Li transforms into conventional FCC metal when $\rho/\rho_0$<0.75. The enthalpies of defective FCC-Li at different densities are shown in Fig. 5.

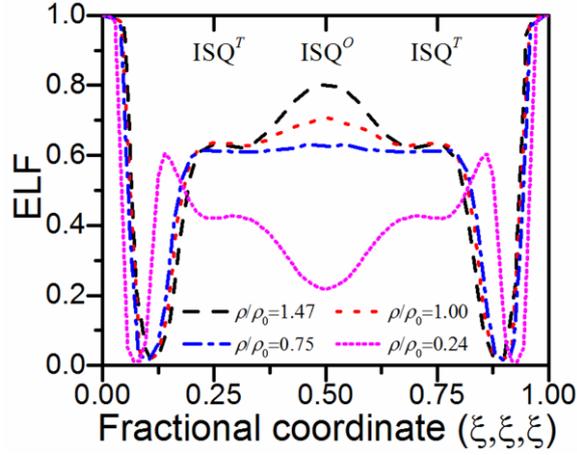

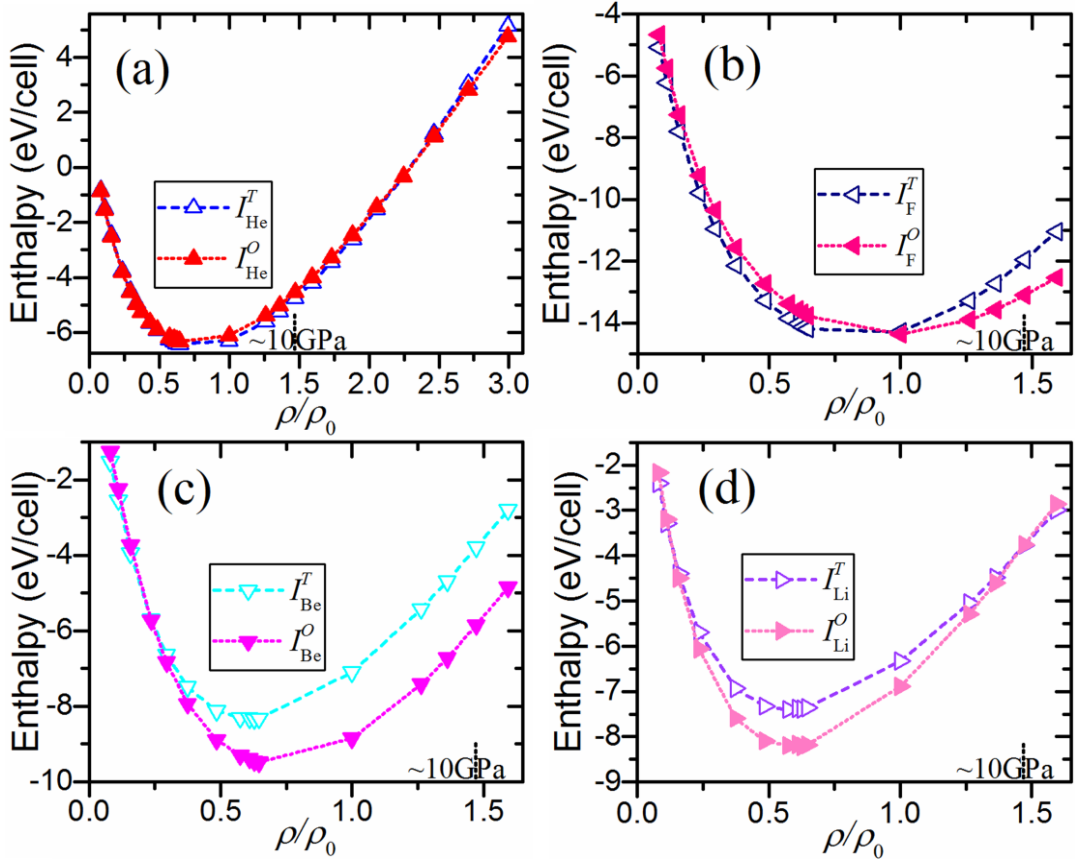

**Figure 4.** (Color online) ELF values along the body diagonal in perfect FCC-Li (see Fig. 1) with various material densities. The interstitial sites are also denoted.

**Figure 5.** (Color online) Calculated enthalpy of Li$_4$X (X=He, F, Be, and self-interstitial Li) with various material densities.

As shown in Fig. 5, the preferred occupation site in 1×1×1 cubic cell for inertial gas impurity He undergoes a transition from the $O$ site to the $T$ site, which is similar to that in the 3×3×3 supercell. The small difference in enthalpy between the two occupation cases should be attributed to the strong interaction between the ISQs and the He and matrix Li atoms, arising from the high density (1/12) of defects. For the high-electronegativity impurity F, the preferred occupation site will become the *T* site in the expanded state, which also can be seen in Fig. 3. This transition arises from the strong electrostatic interactions between the matrix Li atoms and the interstitial F atom, as well as the attractive interaction with the localized electrons near the *T* sites (see Fig. 4). For the low-electronegativity impurity Be and the self-interstitial impurity Li (also can be regarded as a low-electronegativity impurity), the preferred occupation site in the expanded state is the *O* site, the same as in typical FCC metals (as listed in Table II). These results show that this interesting occupation behavior is a unique phenomenon of electrides, arising from the strong interaction between the localized interstitial electrons and the impurity atoms.

**3.2 Tuning on electron localization**

**3.2.1 Charge state of ISQ, defects, and lattice atoms**

**(a) Inertial gas impurity: He**

The switch in the preferred defect sites of the He impurity illustrates the prominent influence on the interstitial occupation of the localized interstitial electrons. The electronic structures of these two systems ($I_{\text{He}}^{O}$ and $I_{\text{He}}^{T}$) are therefore analyzed and compared below. It can be seen that for the interstitial defect $I_{\text{He}}^{T}$ in FCC-Li at 10 GPa (Fig. 6a), the Bader charge of the site ISQ$^O$(NN) changes from a value of 0.64$e$ in perfect FCC-Li to a value of 0.15$e$. For $I_{\text{He}}^{O}$ at 10 GPa (Fig. 6b), the Bader charge of the nearby sites ISQ$^O$(NN) increases to a value of 0.92$e$ ($\Delta Q$=+0.27$e$). This change is a consequence of the strong repulsion between the He impurity and the localized anionic electrons. As shown in Fig. 1, in perfect FCC-Li, the *O* site is occupied by localized electrons and forms anionic ISQ, with a Bader charge of 0.64$e$. However, the occupation of the inertial gas atom He at this *O* site repels the localized electrons, and forces them to transfer to the nearest *O* sites (i.e. enhance the ISQ$^O$(NN)). At the same time, the Bader charge of the nearby *T* site (i.e. ISQ$^T$(NN)) changes from 0.06$e$ to 0.00$e$. In contrast, in the case of $I_{\text{He}}^{T}$, since the number of localized electrons at the *T* site (Fig. 6c) is negligible (with a Bader charge of 0.06$e$), the introduction of the He impurity at the *T* site repels the localized electrons in the ISQ$^O$(NN), meaning that they have a smaller Bader charge, whereas the charge on the ISQ$^T$(NNN) sites increases from 0.06$e$ to 0.19$e$. In other words, doping electrides with the inertial gas impurity He will effectively tune the localized electrons and anionic ISQs by redistributing them due to the repulsive interaction between the localized electrons and the He atom.

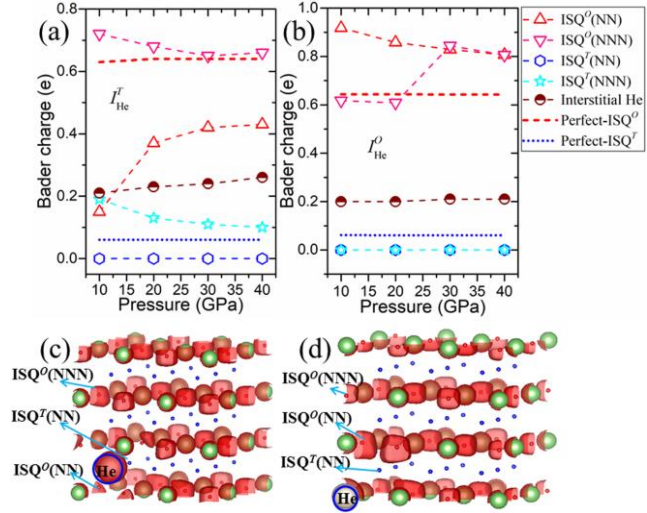

**Figure 6. (Color online) Variation of the Bader charges of the matrix Li atoms, impurity He atom, anionic ISQ$^O$ and ISQ$^T$ nearby the defect of (a) $I_{He}^T$ and (b) $I_{He}^O$. (c) and (d) show the ELF (isosurface=0.75) around these two defects at 10 GPa, respectively. Impurity He atom has a charge state of He$^{0.2-}$.**

It is worth noting that there is a sharp change in the charge states of the ISQ$^O$(NNN) site ($\Delta Q$=+0.20$e$) in $I_{He}^O$ and the ISQ$^O$(NN) site ($\Delta Q$=+0.22$e$) in $I_{He}^T$. This change should be attributed to the redistribution of the anionic electrons. Because the total electrons in the four ISQ$^O$(NN), twelve ISQ$^O$(NNN), and twelve ISQ$^T$(NNN) sites in $I_{He}^T$ and the twelve ISQ$^O$(NN) and six ISQ$^O$(NNN) sites in $I_{He}^O$ are all constant over the whole pressure range studied here (~11$e$ and 15$e$, respectively). This indicates that the interplay between the anionic ISQs, which is driven by pressure, alters the relative stability of these two systems. In other words, the redistribution of localized electrons plays an important role in the stability of point defects in the electride.

**(b) High-electronegativity impurity: F**

For the interstitial defect $I_F^O$, as shown in Fig. 7a, the variation in the Bader charge with pressure is small. It is interesting to note that the Bader charge of ISQ$^O$(NNN) decreases from 0.64$e$ to 0.38$e$, but that of ISQ$^O$(NN) undergoes a negligible change (from 0.64$e$ to 0.67$e$) by the F impurity at 10 GPa compared with the perfect FCC-Li. The Bader charge of the F impurity is ~1.1$e$, and this remains constant over the whole pressure range studied here.

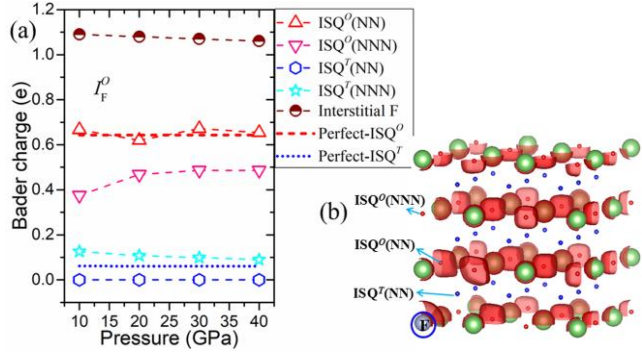

**Figure 7. (Color online) (a) Pressure dependences of the Bader charge of the matrix Li atoms, impurity F atom, anionic ISQ$^O$ and ISQ$^T$, respectively. (b) ELF (isosurface=0.75) around the defective $I_F^O$ site (10 GPa). The impurity F atom has a charge state of F$^{1.1-}$.**

It is evident that due to the high electronegativity, the F atom bonds strongly to the matrix and has a deep formation energy. The Bader charge on the F atom is close to its nominal chemical valance state, and it could be speculated that the F impurity forms an ionic bond with one nearby Li atom. However, the $I_F^O$-Li(NN) distance is much shorter than both the length of the ionic Li-F bond in crystalline LiF (see Fig. S17) and the sum of the ionic radii of Li$^{+1}$ (0.76 Å) and F$^{-1}$ (1.33 Å). A detailed inspection reveals that there is no charge transfer from Li(NN) to $I_F^O$, contrary to expectations. The electrons absorbed by the F atom mainly come from the localized electrons in the $O$ site it occupied, as well as from ISQ$^O$(NNN). It is interesting that the ISQ$^O$(NN) sites retain maintains their localized electrons, and only the shape is highly distorted. The charge redistribution among F, Li(NN), ISQ$^O$(NN), and ISQ$^O$(NNN) is unexpected. We note the fact that Li(NN)-ISQ$^O$(NN) and Li(NN)-ISQ$^O$(NNN) have the same distance, whereas ISQ$^O$(NN) is closer to $I_F^O$ than ISQ$^O$(NNN). The charge transfer from ISQ$^O$(NNN) to $I_F^O$ must therefore take place via a two-step process: (i) electrons are absorbed from Li(NN) to $I_F^O$; and (ii) strong attraction causes electrons to move from ISQ$^O$(NNN) to Li(NN), to compensate for the charge loss. The sequence of these processes is currently unknown, and they may occur simultaneously. ISQ$^O$(NN) retains its localized electrons due to the attraction from the F atom. The same attraction means that the ISQ$^T$(NNN) sites have small numbers of localized excess electrons.

**(c) Low-electronegativity impurity: Be**

Different from the inertial gas atom He and the high electronegativity atom F, the interstitial $I_{Be}^{T}$ has a high Bader charge of ~3.8$e$, meaning that all electrons in the ISQ$^O$(NN) sites are absorbed by the Be impurity, causing the Bader charge of ISQ$^O$(NN) vanish. In contrast, the strong electron localization at ISQ$^O$(NNN) remains. In comparison with $I_{F}^{O}$, the difference between the case of $I_{F}^{O}$ and $I_{Be}^{T}$ mainly lies in the fact that after absorbing one electron, the F atom is saturated and forms a spherical electron cloud (closed shell), whereas there is no saturation in Be and the resultant valence electrons form tetrahedral 'multicenter bonds'.

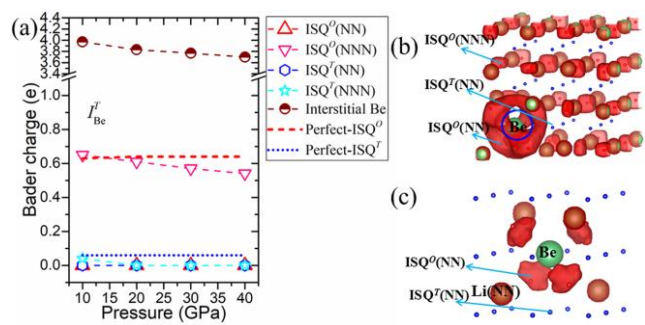

**Figure 8. (Color online) (a) Pressure dependences of the Bader charge of the matrix Li atoms, impurity Be atom, anionic ISQ$^O$ and ISQ$^T$, respectively. (b) ELF (isosurface=0.75) around the defective $I_{Be}^{T}$ site (10 GPa). (c) Local electron localization characteristics of $I_{Be}^{T}$ with ELF isosurface=0.80. The impurity Be atom has a charge state of Be$^{4-}$.**

As shown in Fig. 8c, the local bonding characteristics of this defect are similar to those of the covalent compound diamond (see Fig. S18); in other words, there can be $sp^3$ bonds surrounding the Be atom. However, the electronic density of states (DOS) does not show a clear orbital hybridization feature, as shown in Fig. 11c. Detailed inspection reveals that the radius of the sphere occupied by the excess electrons is 2.35 Å at 10 GPa, which is larger than the distance of $I_{Be}^{T}$-Li(NN) (2.16 Å) and $I_{Be}^{T}$-ISQ$^O$(NN) (1.64 Å) at the same pressure. Although it decreases slightly at 40 GPa (2.02 Å), it is still larger than $I_{Be}^{T}$-Li(NN) (1.99 Å) and $I_{Be}^{T}$ - ISQ$^{O}$(NN) (1.44 Å). That is to say, the electrons of the Be impurity atom occupy a large volume containing both the Li(NN) and ISQ$^{O}$(NN) sites. Hence, the '*sp*³-like' character of the ELF shape of this defect is the consequence of the spatial symmetry of the matrix lattice. In other words, the strong repulsion between $I_{Be}^{T}$ and Li(NN) leads to the vanishing of the electrons between them, meaning that the ELF shows a '*sp*³-like' character that is very different from conventional multicenter covalent bonds[67,68].

**(d) Self-interstitial: Li**

As shown in Fig. 9a, a striking feature for $I_{Li}^{T}$ is that the Bader charge values of ISQ$^{O}$(NN) increase to a value of about 0.8*e*. The values for the electrons in ISQ$^{T}$(NN) also increase from ~0.06*e* to ~0.2*e*. These are very different from the system $I_{Be}^{T}$, although they have a similar ELF shape around the defect. From Fig. 9b and Fig. 11, we can see that there is obvious orbital hybridization between $I_{Li}^{T}$ and Li(NN). It should also be pointed out that the average radius of the sphere occupied by the electrons assigned to the Li impurity atom (1.01 and 0.95 Å at 10 and 40 GPa) by the Bader charge analysis is shorter than the distance of $I_{Li}^{T}$-Li(NN) (2.19 and 1.93 Å) and $I_{Li}^{T}$-ISQ$^{O}$(NN) (1.65 and 1.45 Å) at the same pressure. The electron localization characteristics of this defect therefore result from the formation of *sp*³ bonds.

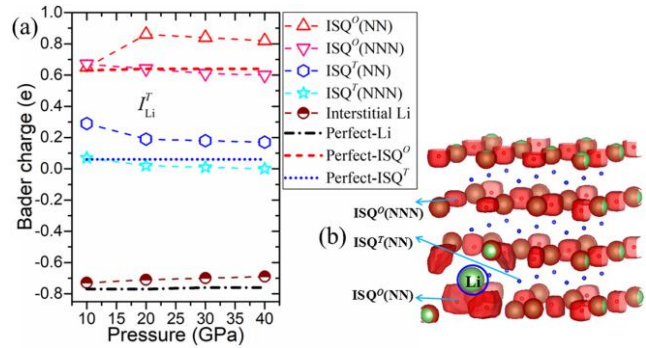

**Figure 9. (Color online) (a) Pressure dependences of the Bader charge of the matrix Li atoms, self-interstitial Li atom, anionic ISQ$^{O}$, and ISQ$^{T}$, respectively. (b) ELF (isosurface=0.75) around the defective $I_{Li}^{T}$ site (10 GPa). The impurity Li atom has a charge state of Li$^{0.7+}$.**

It should be noted that the Bader charges of ISQ$^O$(NN) and ISQ$^T$(NN) undergo a sharp change between 10 and 20 GPa. The increase in the charge for ISQ$^O$(NN) ($\Delta Q$=+0.21$e$) clearly comes mainly from ISQ$^T$(NN) ($\Delta Q$=-0.10$e$). Interstitial Li has a similar Bader charge to the matrix Li atoms at the pressures studied here. In addition, from Fig. 9b we can conclude that doping an interstitial Li into the $T$ site greatly enhances the localization of the electrons of the nearby ISQs. An analysis of the volume enclosed by the ELF (isosurface=0.7) shows that it is increased by 10% compared with perfect FCC-Li at 10 GPa. These results reveal the complex interaction between the Li atoms and the ISQs, as well as the strong interplay between ISQ$^O$(NN) and ISQ$^T$(NN), driven by pressure.

**3.2.2 ELF shape around the defect**

Interstitial point defects not only alter the charge state, but also change the ELF shape around the defective sites compared with perfect FCC-Li. As shown in Figs. 6-9 (as well as Figs. S20 and S21 in SI), the ELF shape of ISQ$^O$(NN) in the defective system is significantly changed by the interstitial atom. The ELF shape of a faraway site, such as ISQ$^O$(NNN) in $I_{\text{Be}}^T$, can also be modified. This indicates the strong interaction among atoms/ISQs. This complex interplay induced by interstitial defects at high pressure is different from the pure pressure effects. As shown by Chen et al[12], they predicted that the alkali metal Li and Na would form an electride compound via the anionic electrons locating at the interstitial sites driven by compression. Also, Miao and Hoffmann[69] showed that an ISQ might form what appear to be covalent bonds with neighboring ISQs or matrix atoms. Hence, the introduction of defects into an electride can offer new ways to tuning or manipulating the electron localization, in addition to the traditional approach of simple compression.

**3.2.3 Energy bands and DOS**

Figure 10 shows the band structure near the Fermi surface (-2~2 eV) of systems with a low relative energy of defect formation (i.e. $I_{\text{He}}^O$, $I_{\text{He}}^T$, $I_{\text{Be}}^T$, $I_{\text{F}}^O$, and $I_{\text{Li}}^T$) at 10 GPa. The detailed data can be found in SI (Fig. S24). We can see that the influence on the band structure mainly manifests as splitting and flattening of the bands. More specifically, He, F, and Be impurity atoms all have an influence, both above and below the Fermi surface. However, the modification of the energy bands near and above the Fermi surface, induced by the inertial gas impurity atom He and the high-electronegativity interstitial F atom, is more obvious than that below the Fermi surface. In contrast, the interstitial Li impurity only has an obvious influence above the Fermi surface. In addition, the He and F impurity atoms will induce the appearance of new energy bands near the Fermi surface at the Γ point, as circled in Fig. 10a and 10b. In particular, these new bands in the $I_{He}^{T}$ system exhibit stronger electron localization feature, indicating that the introduction of interstitial atoms enhances the electron localization. Band degeneracy at here may be attributed to the change in local crystal symmetry induced by the impurity atoms.

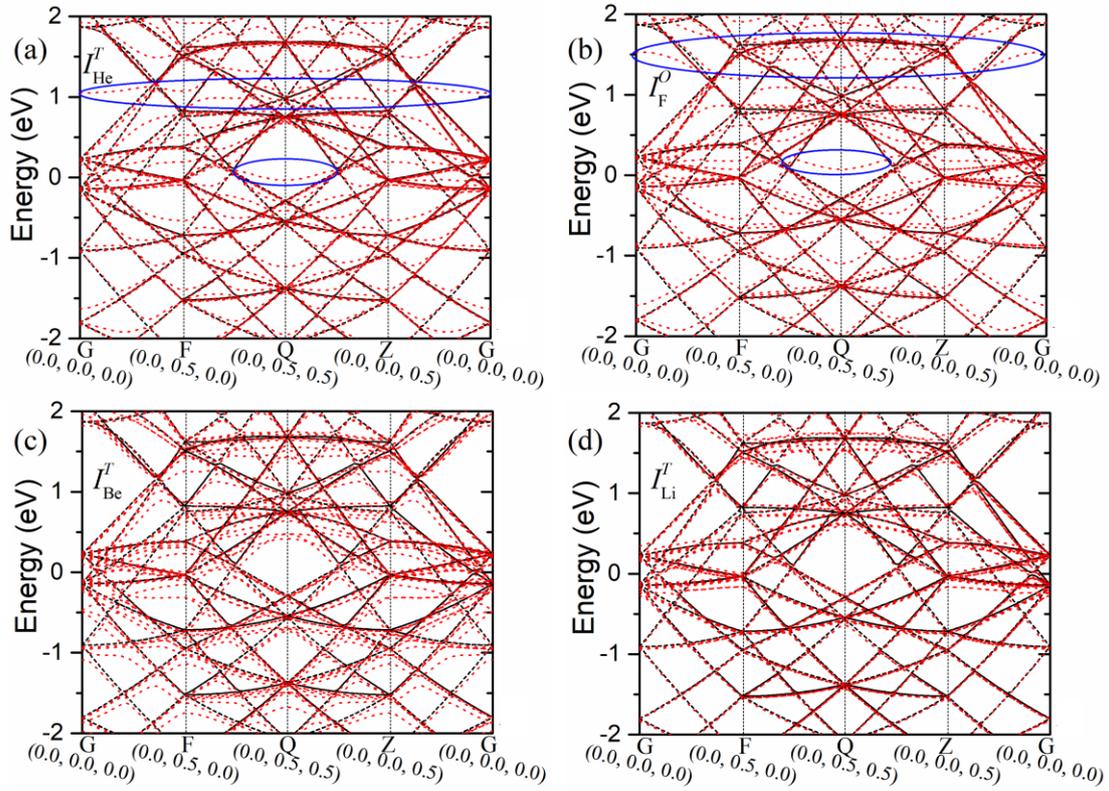

**Figure 10. (Color online) Band structures of the relative stable defective systems at 10 GPa. Red lines represent the defective systems, and black lines represent the perfect FCC-Li.**

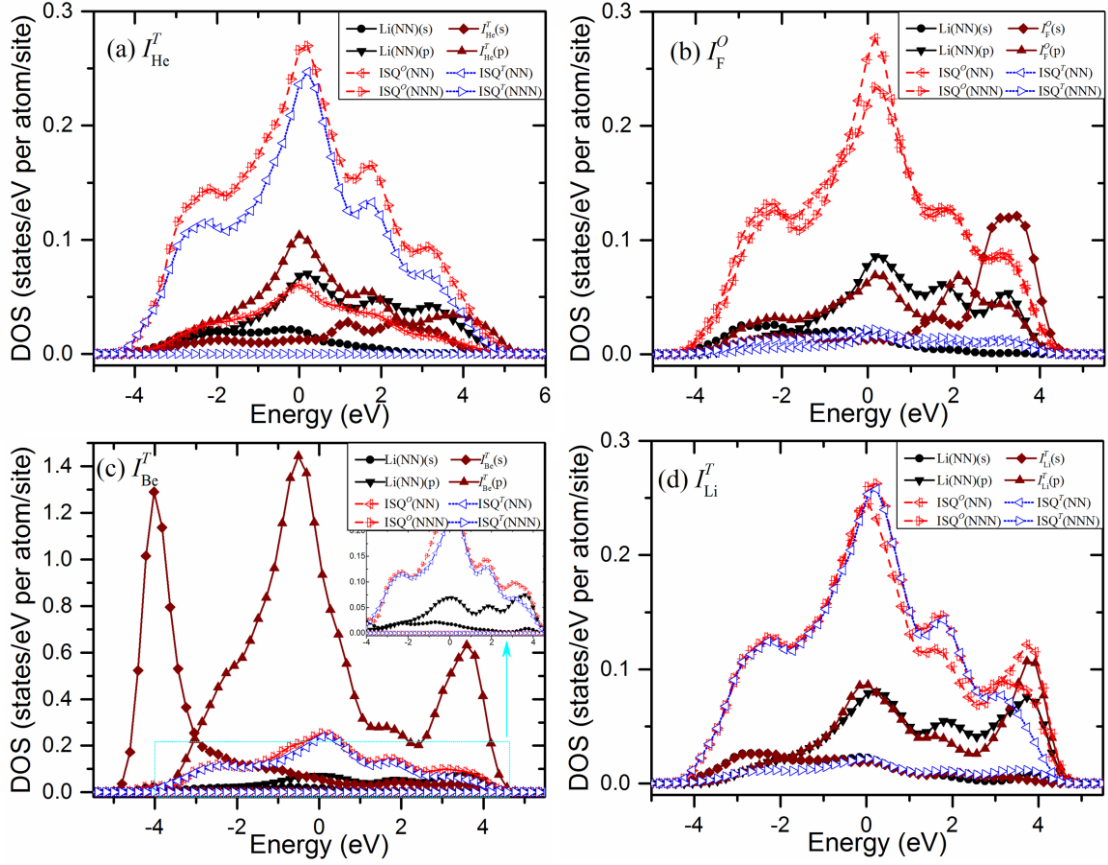

**Figure 11. (Color online)** Calculated local projected DOS of defective FCC-Li with $I_{He}^{T}$, $I_{F}^{O}$, $I_{Be}^{T}$, and $I_{Li}^{T}$ impurity in a 3×3×3 cubic FCC-Li supercell at 10 GPa.

Figure 11 shows the local projected DOS for the four defective systems. Unlike the energy bands, the influence on the total DOS (see Fig. S22 in the SI) is negligible due to the low density of defects (1/324); however, the influence on the local DOS is substantial. As shown in Fig. 11 (as well as Fig. S23 in SI), these four types of impurity atoms all make a contribution to the DOS near to the Fermi surface. In particular, the *p*-electron of Be has the largest DOS when compared to the impurity atoms He, F, and Li. This indicates that more excess electrons are absorbed by Be atom, a result that is in agreement with those of the aforementioned Bader charge analysis.

### 3.3 The highest negative valence state record: Be impurity

The valence state reflects the number of electrons that can be present in the outer shell of a given element. These valence electrons determine the physical and chemical properties of the material. It is interesting to note that the low-electronegativity impurity atom Be can form a nominal valence state of $Be^{8-}$ when it occupies the *O* site, as shown in Fig. 12, which is the highest negative valence recorded any conditions and violates the octet rule in chemistry.

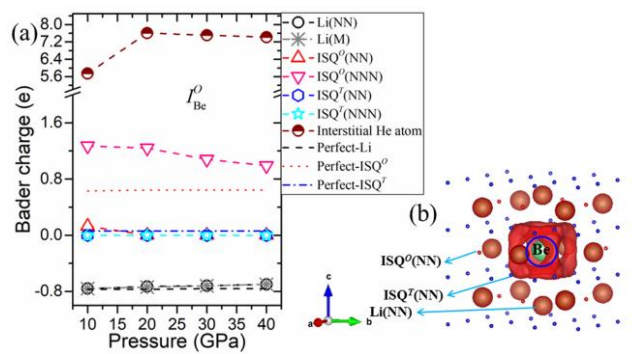

**Figure 12. (Color online) (a) Pressure dependences of the Bader charge of the matrix Li atoms, interstitial Be atom, anionic ISQ$^O$, and ISQ$^T$, respectively. (b) ELF (isosurface=0.75) around the defective $I_{\text{Be}}^{O}$ site (10 GPa). The impurity atom has a remarkable nominal valence state of -8 for Be with an electron configuration of 1$s^2$2$s^2$2$p^6$3$s^2$.**

To explore the source of the electrons for Be impurity, we carried out a detailed analysis of the charge density distribution. As shown in Fig. 13, the Bader partitioning of the impurity atom Be contains the ISQ$^T$(NN) sites but not the ISQ$^O$(NN) sites. This is agreement with the result that the Bader charge of ISQ$^T$(NN) is zero, and that of ISQ$^O$(NN) is 0.13$e$ (see Fig. 12a). A further inspection of the Bader charge of all the atoms and ISQs (see Table S12) reveals that the electrons mainly come from the nearby ISQs. Because the net change of electrons for the twelve ISQ$^O$(NN), six ISQ$^O$(NNN), twenty-four ISQ$^O$(3NN), eight ISQ$^T$(NN), and twenty-four ISQ$^T$(NNN) sites are nearly the same with the charge state of Be.

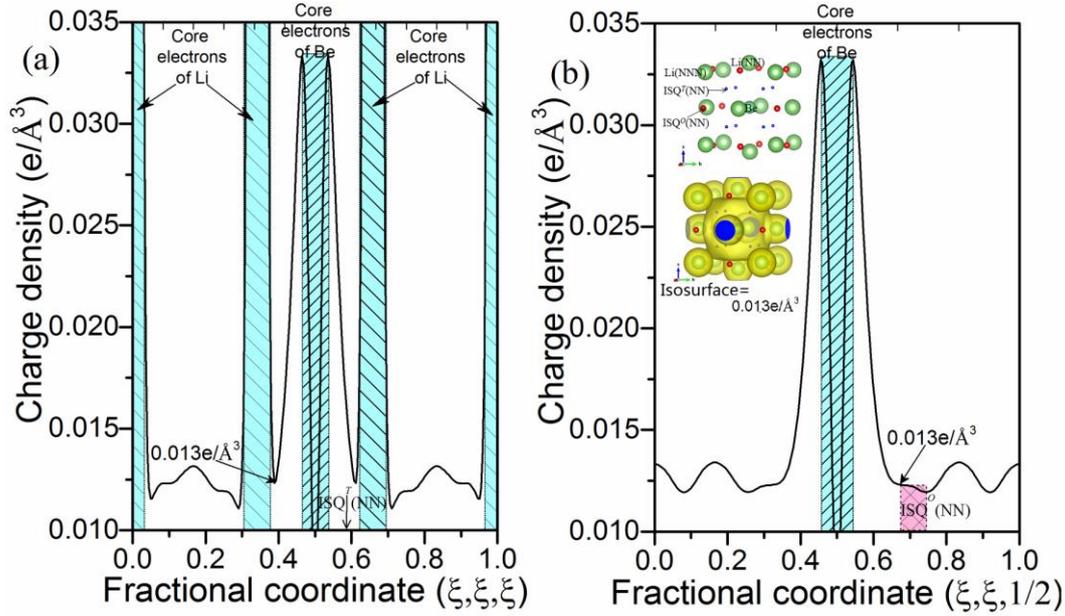

**Figure 13. (Color online) Charge density along the (a) body diagonal and (b) face diagonal in system $I_{Be}^{O}$ at 10 GPa. Insets in (b) show the 3-D charge density around the defective $I_{Be}^{O}$ site (isosurface=0.013 $e$/Å). The sites of atoms and ISQs are also denoted.**

The mechanism by which this remarkable negative valence state forms is as yet unclear, but the results are reliable. Our further calculations verify this conclusion, and show that impurity Mg, Al, and Fe can also form an approximate nominal valence state of $Mg^{8-}$, $Fe^{8-}$, and $Al^{7-}$ (also violates the octet rule) when they occupy the same site at 10 GPa, respectively. This suggests that doping impurity elements into electrides is an efficient way of achieving a very high negative charge state of an atom, which might greatly expand the range of applications.

## IV. Conclusion

Dense FCC-Li was employed as an archetype to investigate the interplay and interaction between the (self-)interstitial impurity atoms (He, F, Be, and Li) and the matrix Li atoms and anionic ISQs in electrides based on the first-principles calculations. The calculated enthalpy of defect formation indicates that it is more favorable for (self-) interstitial atoms (He, Be, Li) to occupy the *T* site in dense FCC-Li. More detailed analyses reveal that this abnormal interstitial occupation behavior is unique in electrides, and is induced by the strong interactions between the ISQs and impurities, contrary to the common wisdom. A Bader charge analysis, an ELF topology analysis, and an energy band and DOS analysis indicate that all of these impurity atoms can enhance the localization of electrons. At the same time, these results reveal a very strong coupling among interstitial impurity atoms, the ISQs, and the matrix Li atoms near to the defects. Which verify the possibility to manipulate the localization characteristics of electrons in electrides and illustrate an efficient route to enrich their variants for wide applications. In particular, we found a remarkable charge state of Be ($Be^{8-}$) when it occupies an $O$ site, which is the highest discovered so far. The fascinating electronic behavior revealed here demonstrates the rich chemistry and physics of this emerging material, as well as the wide range of potential applications. This work also lays a solid foundation for further research on this promising material.

**Conflicts of interest**

The authors declare no competing financial interest.

**Acknowledgements**

This work was supported by the National Natural Science Foundation of China under Grant No. 11672274, the NSAF under Grant No. U1730248, the Science challenge Project under Grant No. TZ2016001, the CAEP Research Project CX2019002. Support from the National Natural Science Foundation of China (11704163, 11804131), China Postdoctoral Science Foundation (2017M623064), and Natural Science Foundation of Jiangxi Province of China (20181BAB211007) are also acknowledged. Part of the computation was performed using the supercomputer at the Center for Computational Materials Science (CCMS) of the Institute for Material Research at Tohoku University, Japan.

**Supporting Information**

Results of convergence test; Detailed data of the interaction between defects and matrix atoms/ISQs: differential ELF between the defective and perfect FCC-Li, radius of the area influenced by interstitial point defect, distance between point defect and matrix atoms/ISQs, and equation of state; Detailed method and results of the Bader charge analysis; Distance of Li-F in crystalline LiF and defective FCC-Li; Example of mislabeled electrides; ELF, electronic DOS and band structure; Results of high defects density.

For Table of Contents Only

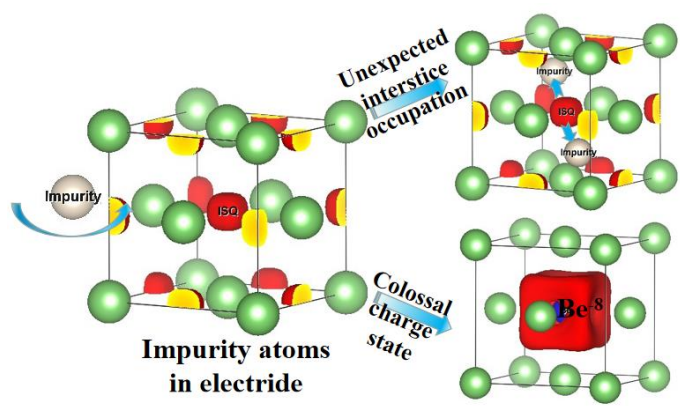

# Supporting Information

# Interplay of anionic quasi-atom and interstitial point defects in electrides: abnormal interstice occupation and colossal charge state of point defects in dense FCC-lithium

Leilei Zhang,[†] Qiang Wu,[†] Shourui Li,[†] Yi Sun,[†] Xiaozhen Yan,[†, ‡] Ying Chen,[+] and Hua Y. Geng[*, †, §]

[†]*National Key Laboratory of Shock Wave and Detonation Physics, Institute of Fluid Physics, CAEP, P.O. Box 919-102, Mianyang, Sichuan 621900, P.R. China*

[‡]*Jiangxi University of Science and Technology, Ganzhou, Jiangxi 341000, P.R. China*

[+]*Fracture and Reliability Research Institute, School of Engineering, Tohoku University 6-6-01 Aramakiaoba, Aoba-ku, Sendai, 980-8579, Japan*

[§]*Center for Applied Physics and Technology, HEDPS, and College of Engineering, Peking University, Beijing 100871, P.R. China*

## Table of Contents

[*] To whom correspondence should be addressed. E-mail: s102genghy@caep.cn

## S1. Results of convergence test

The convergence of total energy with respect to the energy cutoff of the plane wave expansion, supercell size, and k-point meshes are tested carefully. The results are shown in Figs. S1-S3. We can see that the total energy reaches well convergence when the energy cutoff is larger than or equal 850 eV, the long-rang interaction of the interstitial point defect is shorter than the length of three cubic FCC-Li cell, and the total energy can reach convergence when the MP mesh is larger than 2×2×2. Therefore, energy cutoff of 900 eV, supercell size of 3×3×3, MP mesh of 5×5×5 using in this work are sufficient to ensure the convergence of the results.

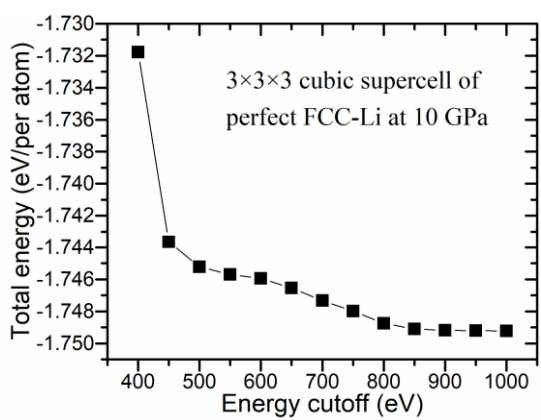

**Figure S1.** Total energy convergence with respect to the energy cutoff for a 3×3×3 cubic FCC-Li supercell calculated by DFT-PBE method.

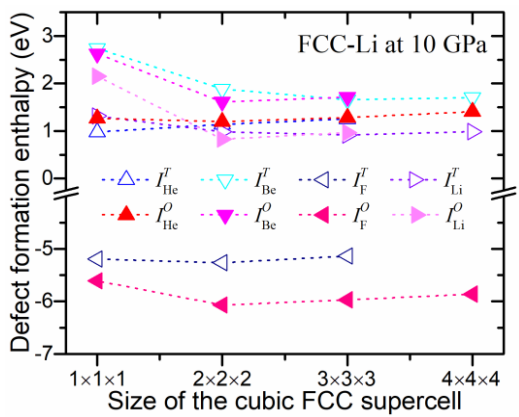

**Figure S2.** Convergence of the calculated formation enthalpy (Δ$H$) of point defects with respect to the supercell size. The Δ$H$ of a defect of Li$m$X ($X$=He, F, Be, and Li) are defined as Δ$H$(Li$m$X)=$H$(Li$m$X)-$mH$(Li)-$H$(X), where $H$(Li$m$X) is the enthalpy of the defective system Li$m$X; $H$(Li) and $H$(X) are the enthalpy per atom of the most favored structure of Li and $X$ at given pressure, respectively. $I^O_{He}$ ($I^T_{He}$), $I^O_{Be}$ ($I^T_{Be}$), $I^O_F$ ($I^T_F$), and $I^O_{Li}$ ($I^T_{Li}$) represent interstitial point defect (He, Be, F, and Li) occupying the $O$($T$)site, respectively.

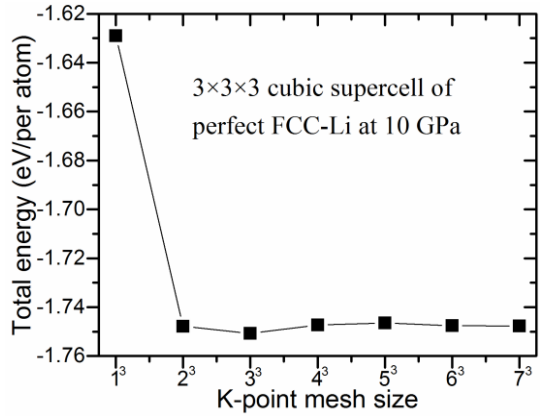

**Figure S3.** Convergence of the calculated total energy per atom of a 3×3×3 cubic supercell of perfect FCC-Li at 10 GPa with respect to the k-point (MP) meshes size.

## S2. Influenced area of interstitial point defects

### S2.1. Tetrahedral interstitial point defects: He, Be, and F

Figure. S4 shows the differential ELF between the defective and perfect FCC-Li, as well as the atomic positions of the defect and matrix atoms/ISQs. The radius ($r$) of the influenced areas are listed in Table S1, which are compared with the lattice parameter ($a$) of the 3×3×3 cubic FCC supercell. The results indicate that interstitial impurity atoms mainly have an influence range to the NN-, NNN-, and 3NN-Li atoms. The change of the distance ($\Delta d$) between the impurity atom (noted as $X$) and the Li(4NN) is very small compared with perfect FCC-Li (see Table S2 and Fig. S5).

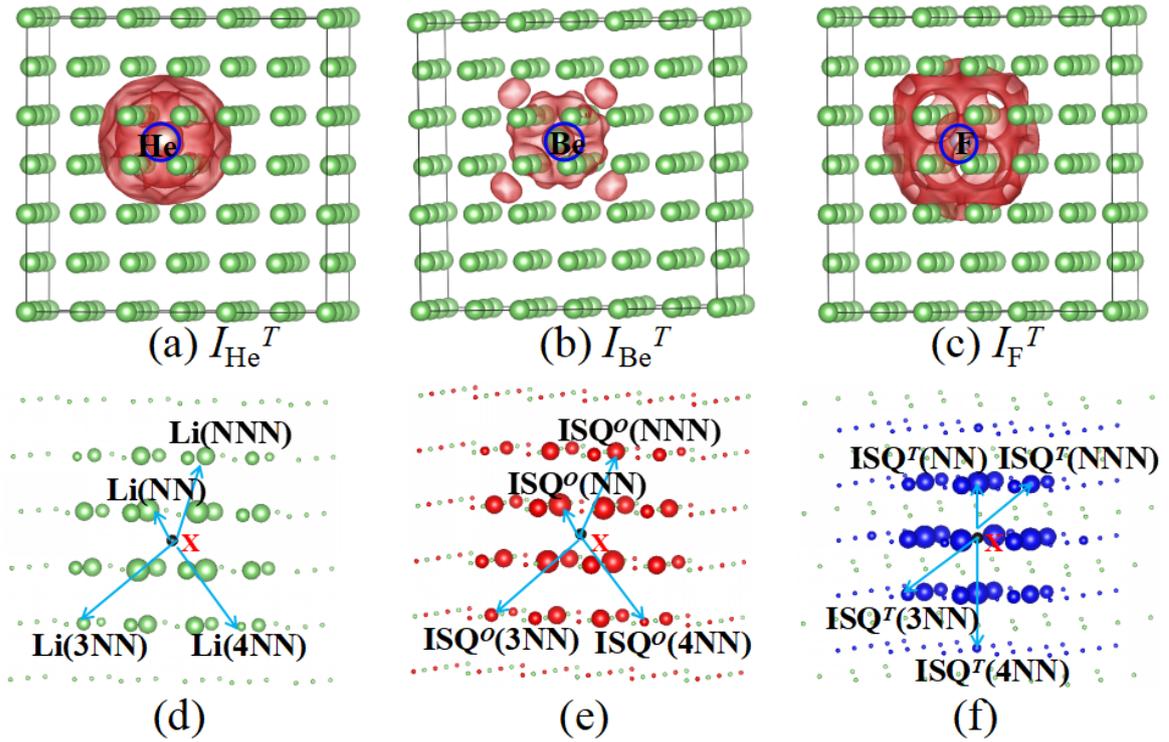

**Figure S4.** The differential ELF (isosurface=0.05) between the defective and perfect FCC-Li (a-c) and atomic positions of the defect and matrix atoms/ISQs (d-f) at 10 GPa, respectively. The point defect sites are marked with a blue circle in (a)-(c) and denoted as $X$ in (d)-(f).

**Table S1.** Radius ($r$, Å) of the area influenced by the tetrahedral interstitial point defect at different pressures (GPa) compared with the corresponding lattice parameter $a$ of a 3×3×3 cubic supercell.

| Pressure | $I_{He}^T$ | | | $I_{Be}^T$ | | | $I_F^T$ | | |
|---|---|---|---|---|---|---|---|---|---|
| | $a$ | $r$ | $a/r$ | $a$ | $r$ | $a/r$ | $a$ | $r$ | $a/r$ |
| 10 | 11.41 | 2.45 | 0.21 | 11.40 | 4.38 | 0.38 | 11.41 | 3.24 | 0.28 |
| 40 | 10.01 | 2.09 | 0.21 | 10.01 | 3.84 | 0.38 | 10.02 | 3.08 | 0.31 |

**Table S2.** The distance ($d$, Å) between the interstitial point defect occupying the $T$ site and the matrix atoms/ISQs at different pressures (GPa). $\Delta d = d - d_0$, where $d_0$ is the according

S5

distance of perfect FCC-Li at the same pressure.

| X | $I^T_{He}$ | | | | $I^T_{Be}$ | | | | $I^T_F$ | | | |
|---|---|---|---|---|---|---|---|---|---|---|---|---|
| Pressure | 10 | | 40 | | 10 | | 40 | | 10 | | 40 | |
| Distance | d | Δd | d | Δd | d | Δd | d | Δd | d | Δd | d | Δd |
| X-Li(NN) | 1.75 | 0.10 | 1.52 | 0.08 | 2.16 | 0.51 | 1.99 | 0.55 | 1.74 | 0.09 | 1.65 | 0..21 |
| X-Li(NNN) | 3.19 | 0.03 | 2.80 | 0.04 | 3.02 | -0.14 | 2.64 | -0.12 | 3.24 | 0.08 | 2.82 | 0.06 |
| X-Li(3NN) | 4.16 | 0.01 | 3.64 | 0.01 | 4.30 | 0.15 | 3.82 | 0.19 | 4.16 | 0.01 | 3.70 | 0.07 |
| X-Li(4NN) | 4.95 | 0.00 | 4.34 | 0.01 | 4.89 | -0.06 | 4.29 | -0.04 | 4.95 | 0.00 | 4.33 | 0.00 |
| X-ISQ$^O$(NN) | 1.65 | 0.00 | 1.44 | 0.00 | 1.64 | -0.01 | 1.44 | 0.00 | 1.65 | 0.00 | 1.45 | 0.01 |
| X-ISQ$^O$(NNN) | 3.15 | -0.01 | 2.77 | 0.01 | 3.15 | -0.01 | 2.77 | 0.01 | 3.15 | -0.01 | 2.77 | 0.01 |
| X-ISQ$^O$(3NN) | 4.14 | -0.01 | 3.64 | 0.01 | 4.14 | -0.01 | 3.64 | 0.01 | 4.15 | 0.00 | 3.64 | 0.01 |
| X-ISQ$^O$(4NN) | 4.94 | -0.01 | 4.33 | 0.00 | 4.93 | -0.02 | 4.33 | 0.00 | 4.94 | -0.01 | 4.34 | 0.01 |
| X-ISQ$^T$(NN) | 1.90 | 0.00 | 1.67 | 0.00 | 1.90 | 0.00 | 1.67 | 0.00 | 1.90 | 0.00 | 1.67 | 0.00 |
| X-ISQ$^T$(NNN) | 2.69 | 0.00 | 2.36 | 0.00 | 2.69 | 0.00 | 2.36 | 0.00 | 2.69 | 0.00 | 2.36 | 0.00 |
| X-ISQ$^T$(3NN) | 3.29 | -0.01 | 2.89 | 0.00 | 3.29 | -0.01 | 2.89 | 0.00 | 3.29 | -0.01 | 2.89 | 0.00 |
| X-ISQ$^T$(4NN) | 3.80 | -0.01 | 3.34 | 0.01 | 3.80 | -0.01 | 3.34 | 0.01 | 3.80 | -0.01 | 3.34 | 0.01 |
| X-ISQ$^T$(5NN) | 4.25 | -0.01 | 3.73 | 0.01 | 4.25 | -0.01 | 3.73 | 0.01 | 4.25 | -0.01 | 3.73 | 0.01 |
| X-ISQ$^T$(6NN) | 4.66 | 0.00 | 4.09 | 0.01 | 4.65 | -0.01 | 4.09 | 0.01 | 4.66 | 0.00 | 4.09 | 0.01 |
| X-ISQ$^T$(7NN) | 5.38 | -0.01 | 4.72 | 0.01 | 5.37 | -0.02 | 4.72 | 0.01 | 5.38 | -0.01 | 4.72 | 0.01 |

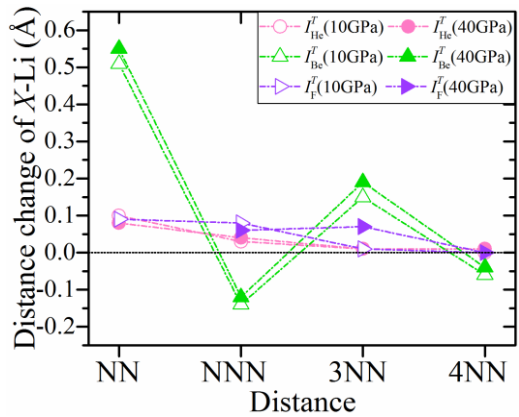

**Figure S5.** Change of the distance between the impurity atom occupying the *T* sites and the nearby Li atoms compared with the perfect FCC-Li at different pressures.

### S2.2. Octahedral interstitial point defects: He, Be, and F

Figure. S6 shows the differential ELF between the defective and perfect FCC-Li, as well as the atomic position of the defect and matrix atoms/ISQs. Impurity Be occupying *O* site ($I^O_{Be}$) has the largest influence on the ELF of the perfect FCC-Li. The influenced area contains mainly the NN, NNN, and 3NN matrix Li atoms and anionic ISQs. The radius (*r*) of the influenced areas, as listed in Table S3, show that the relative distance influenced by the interstitial impurity atoms increases slightly with the increasing pressure, but remains less than half of the supercell lattice parameter $a$. For example, it increases from 36% to 39% when the pressure increased from 10 GPa to 40 GPa for $I_{Be}^{O}$. We also calculated the change of the distance ($\Delta d$) between the impurity atom (noted as $X$) and the nearby Li/ISQs with respect to the perfect FCC-Li at different pressures. The results, as shown in Table S4 and Fig. S7, also indicate that interstitial impurity atoms mainly have an influence on the NN-, NNN-, and 3NN-Li atoms.

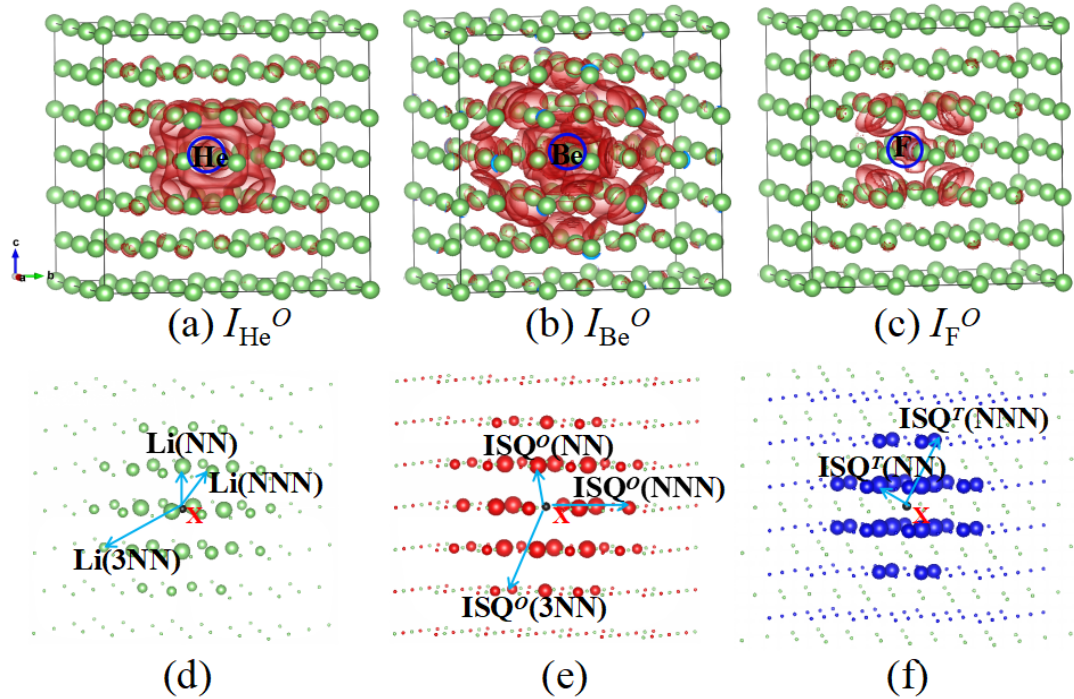

**Figure S6.** The differential ELF (isosurface=0.1) between the defective and perfect FCC-Li (a-c) and the atomic positions of the defect and matrix atoms/ISQs (d-f) at 10 GPa, respectively. The point defect sites are marked with a blue circle in (a)-(c) and denoted as $X$ in (d)-(f).

**Table S3.** Radius ($r$, Å) of the area influenced by the octahedral interstitial point defect at different pressures (GPa) compared with the according lattice parameter $a$ of a 3×3×3 cubic supercell.

| Pressure | $I_{He}^{O}$ | | | $I_{Be}^{O}$ | | | $I_{F}^{O}$ | | |
|---|---|---|---|---|---|---|---|---|---|
| | $a$ | $r$ | $r/a$ | $a$ | $r$ | $r/a$ | $a$ | $r$ | $r/a$ |
| 10 | 11.43 | 2.62 | 0.23 | 11.41 | 4.12 | 0.36 | 11.40 | 2.89 | 0.25 |
| 40 | 10.00 | 2.45 | 0.25 | 10.01 | 3.92 | 0.39 | 10.01 | 2.53 | 0.25 |

**Table S4.** The distance ($d$, Å) between the interstitial point defect occupying the $O$ site and matrix atoms/ISQs at different pressures (GPa). $\Delta d=d-d_0$, where $d_0$ is the according distance

of perfect FCC-Li at the same pressure.

| X | $I_{He}^{O}$ | | | | $I_{Be}^{O}$ | | | | $I_{F}^{O}$ | | | |
|---|---|---|---|---|---|---|---|---|---|---|---|---|
| Pressure | 10 | | 40 | | 10 | | 40 | | 10 | | 40 | |
| Distance | d | Δd | d | Δd | d | Δd | d | Δd | d | Δd | d | Δd |
| X-Li(NN) | 2.02 | 0.12 | 1.73 | 0.06 | 2.21 | 0.31 | 2.02 | 0.35 | 1.87 | -0.03 | 1.70 | 0.03 |
| X-Li(NNN) | 3.34 | 0.04 | 2.96 | 0.07 | 3.13 | -0.17 | 2.72 | -0.17 | 3.37 | 0.07 | 2.95 | 0.06 |
| X-Li(3NN) | 4.26 | 0.00 | 3.71 | -0.01 | 4.34 | 0.08 | 3.82 | 0.10 | 4.23 | -0.03 | 3.74 | 0.02 |
| X-Li(4NN) | 5.72 | 0.01 | 5.00 | 0.00 | 5.70 | 0.01 | 5.01 | 0.01 | 5.70 | -0.01 | 5.01 | 0.01 |
| X-ISQ$^O$(NN) | 2.69 | 0.00 | 2.36 | 0.00 | 2.69 | 0.00 | 2.36 | 0.00 | 2.69 | 0.00 | 2.36 | 0.00 |
| X-ISQ$^O$(NNN) | 3.81 | 0.00 | 3.33 | 0.00 | 3.81 | 0.00 | 3.34 | 0.01 | 3.80 | -0.01 | 3.34 | 0.01 |
| X-ISQ$^O$(3NN) | 4.67 | 0.01 | 4.08 | 0.00 | 4.66 | 0.00 | 4.09 | 0.01 | 4.65 | -0.01 | 4.09 | 0.01 |
| X-ISQ$^T$(NN) | 1.65 | 0.00 | 1.44 | 0.00 | 1.65 | 0.00 | 1.45 | 0.01 | 1.64 | -0.01 | 1.44 | 0.00 |
| X-ISQ$^T$(NNN) | 3.16 | 0.00 | 2.77 | 0.01 | 3.15 | -0.01 | 2.77 | 0.01 | 3.15 | -0.01 | 2.77 | 0.01 |

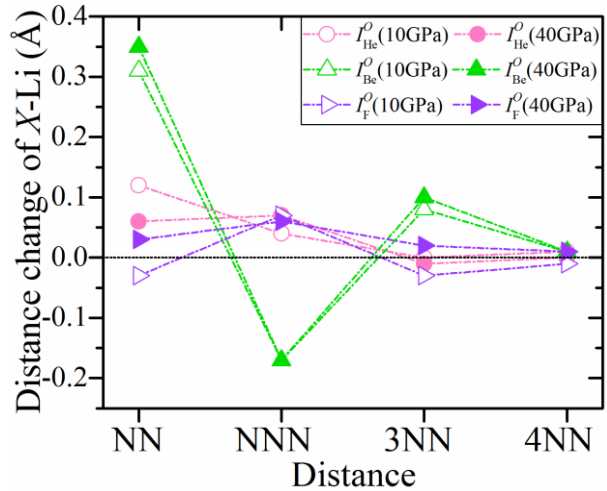

**Figure S7.** Change of the distance between the impurity atom occupying the *O* sites and the nearby Li atoms compared with the perfect FCC-Li at different pressures.

### S2.3. Self-interstitial point defect: Li

Figure. S8 shows the differential ELF between the defective and perfect FCC-Li. The radius (*r*) of the influenced areas are listed in Table S5, which are compared with the lattice parameter (*a*) of the 3×3×3 cubic FCC supercell. The change of the distance (*Δd*) between the impurity atom and the nearby Li/ISQs compared with perfect FCC-Li at different pressures are shown in Table S6 and Fig. S9. These results also indicate that interstitial impurity atoms mainly have an influence range on the NN-, NNN-, and 3NN-Li atoms.

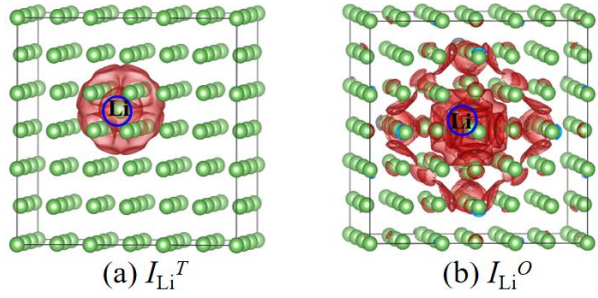

(a) $I_{Li}^{T}$　　　　(b) $I_{Li}^{O}$

**Figure S8.** The differential ELF (isosurface=0.05) between the defective and perfect FCC-Li. The point defect sites are marked with a blue circle.

**Table S5.** Radius ($r$, Å) of the area influenced by the self-interstitial point defect at different pressures (GPa) compared with the correspondingly lattice parameter $a$ of a 3×3×3 cubic supercell.

| Pressure | $I_{Li}^{T}$ | | | $I_{Li}^{O}$ | | |
|---|---|---|---|---|---|---|
| | $a$ | $r$ | $a/r$ | $a$ | $r$ | $a/r$ |
| 10 | 11.42 | 4.36 | 0.38 | 11.42 | 4.30 | 0.38 |
| 40 | 10.02 | 3.97 | 0.40 | 10.03 | 3.90 | 0.39 |

**Table S6.** The distance ($d$, Å) between the self-interstitial point defect occupying the $T/O$ site and matrix atoms/ISQs at different pressures (GPa). $\Delta d = d - d_0$, where $d_0$ is the according distance of perfect FCC-Li at the same pressure.

| X | $I_{Li}^{T}$ | | | | $I_{Li}^{O}$ | | | |
|---|---|---|---|---|---|---|---|---|
| Pressure | 10 | | 40 | | 10 | | 40 | |
| Distance | $d$ | $\Delta d$ | $d$ | $\Delta d$ | $d$ | $\Delta d$ | $d$ | $\Delta d$ |
| $X$-Li(NN) | 2.19 | 0.54 | 1.93 | 0.49 | 2.27 | 0.37 | 2.00 | 0.33 |
| $X$-Li(NNN) | 3.11 | -0.05 | 2.73 | -0.03 | 3.22 | -0.08 | 2.84 | -0.05 |
| $X$-Li(3NN) | 4.31 | 0.16 | 3.79 | 0.16 | 4.34 | 0.08 | 3.80 | 0.08 |
| $X$-Li(4NN) | 4.92 | -0.03 | 4.32 | -0.01 | 5.71 | 0.00 | 5.01 | 0.01 |
| $X$-ISQ$^O$(NN) | 1.65 | 0.00 | 1.45 | 0.01 | 2.69 | 0.00 | 2.36 | 0.00 |
| $X$-ISQ$^O$(NNN) | 3.16 | 0.00 | 2.77 | 0.01 | 3.81 | 0.00 | 3.34 | 0.01 |
| $X$-ISQ$^O$(3NN) | 4.15 | 0.00 | 3.64 | 0.01 | 4.66 | 0.00 | 4.09 | 0.01 |
| $X$-ISQ$^O$(4NN) | 4.95 | 0.00 | 4.34 | 0.01 | - | - | - | - |
| $X$-ISQ$^T$(NN) | 1.90 | 0.00 | 1.67 | 0.00 | 1.65 | 0.00 | 1.45 | 0.01 |
| $X$-ISQ$^T$(NNN) | 2.69 | 0.00 | 2.36 | 0.00 | 3.16 | 0.00 | 2.77 | 0.01 |
| $X$-ISQ$^T$(3NN) | 3.30 | 0.00 | 2.89 | 0.00 | - | - | - | - |
| $X$-ISQ$^T$(4NN) | 3.81 | 0.00 | 3.34 | 0.01 | - | - | - | - |
| $X$-ISQ$^T$(5NN) | 4.26 | 0.00 | 3.73 | 0.01 | - | - | - | - |
| $X$-ISQ$^T$(6NN) | 4.66 | 0.00 | 4.09 | 0.01 | - | - | - | - |
| $X$-ISQ$^T$(7NN) | 5.38 | -0.01 | 4.72 | 0.01 | - | - | - | - |

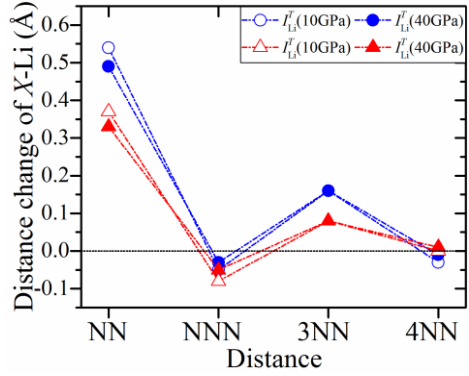

**Figure S9.** Change of the distance between the self-interstitial impurity atom and the nearby Li atoms compared with the perfect FCC-Li at different pressures.

## S3. Equation of state

As discussed above, interstitial impurity atoms have obvious influence on the nearly matrix. This implies that interstitial impurity atoms could change the equation of state, at least locally. The volume variations of the 3×3×3 cubic FCC supercell of these defective systems are shown in Fig. S10(a). We can see that the cell volume does not have obvious increase or decrease compared with the perfect FCC-Li, due to the very low defect density (1/324) in this case. Besides, an 1×1×1 cubic FCC cell is also used to evaluate the volume variation (corresponding to a defect density of 1/12), as shown in Fig. S10(b). The volume of all these 1×1×1 cubic FCC defective systems increase compared with perfect FCC-Li. It is necessary to note that the cell volume has a continuous variation with the pressure for all defective systems. We noticed that $I_{Li}^{O}$ and $I_{Be}^{O}$ have a bigger volume than $I_{Li}^{T}$ and $I_{Be}^{T}$, which is quite counter-intuitive.

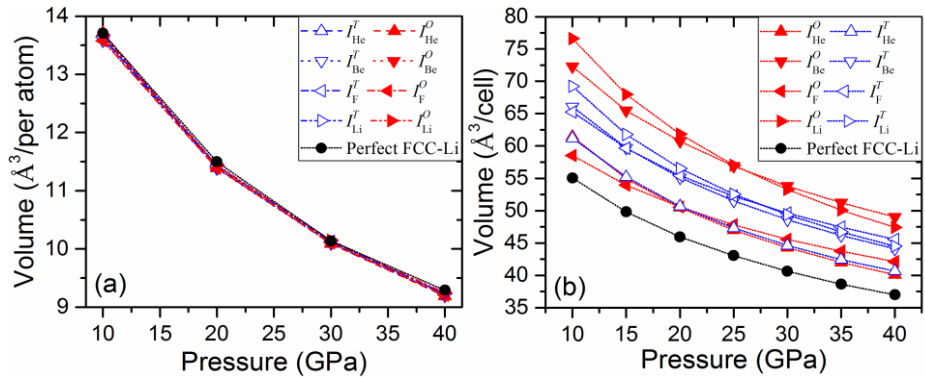

**Figure S10.** Variation of the atomic volume of the 3×3×3 defective FCC-Li supercell (a) and variation of the cell volume of the 1×1×1 defective FCC-Li cubic cell (b) calculated by DFT-PBE method.

## S4. Bader charge analysis

In order to explore the distribution and the variance of Bader charge of the matrix atoms and anionic ISQs surrounding the point defect, we grouped each atom/ISQ according to their distance to the point defect (see Fig. 2 in the main text), and then making statistical analysis within each group. Both the averaged value ($q$) and the standard deviation ($\delta q$) are evaluated. The change with respect to the value of the counterpart in the perfect FCC-Li lattice at the same pressure is also estimated ($\Delta Q$).

For each coordination shell, the averaged Bader charge is estimated by

$$q = \frac{1}{N}\sum_{i=1}^{N} q_i \qquad (S1)$$

The standard deviation is calculated by

$$\delta q = \sqrt{\frac{1}{N}\sum_{i=1}^{N}(q_i - q)^2} \qquad (S2)$$

The deviation from the perfect FCC-Li is calculated by

$$\Delta Q = q - q_0 \qquad (S3)$$

Where $q_i$ are the Bader charges of matrix atoms or the anionic ISQs in the defective lattice, and $q_0$ is the corresponding charge in the perfect FCC-Li lattice at the same pressure.

The results of perfect FCC-Li are listed in Table S7. Results of the defective systems are listed in Tables S8-S15, in which the highlighted data indicate that they have an obvious change compared with perfect FCC-Li.

**Table S7.** The calculated average Bader charge ($q$) of matrix Li atoms and the anionic ISQs in perfect FCC-Li at the given pressures, together with their respective standard deviation ($\delta q$).

| site | 10 GPa | | 20 GPa | | 30 GPa | | 40 GPa | |
|---|---|---|---|---|---|---|---|---|
| | Bader charge (e) | | Bader charge (e) | | Bader charge (e) | | Bader charge (e) | |
| | $q$ | $\delta q$ | $q$ | $\delta q$ | $q$ | $\delta q$ | $q$ | $\delta q$ |
| Li | -0.77 | 0.00 | -0.77 | 0.00 | -0.76 | 0.00 | -0.76 | 0.00 |
| ISQ$^O$ | 0.64 | 0.00 | 0.64 | 0.00 | 0.64 | 0.00 | 0.64 | 0.00 |
| ISQ$^T$ | 0.06 | 0.00 | 0.06 | 0.00 | 0.06 | 0.00 | 0.06 | 0.00 |

### S4.1. Tetrahedral interstitial point defects: He, Be, and F

The detailed Bader charge variation of the matrix Li atoms, impurity atom, anionic ISQ$^O$

and ISQ$^T$ of the defective systems $I_{He}^{T}$, $I_{Be}^{T}$, and $I_{F}^{T}$ are shown in Fig. S11. The specific values are listed in Tables S8-S10.

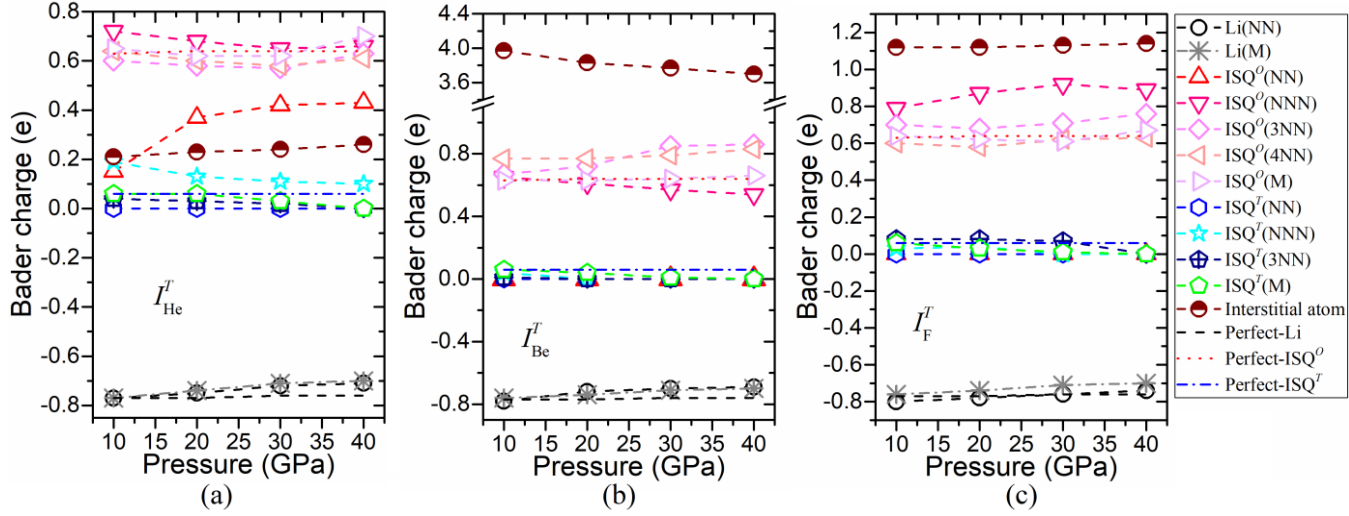

**Figure S11. Variation of the Bader charge of the matrix Li atoms, impurity atom, anionic ISQ$^O$, and ISQ$^T$ of the defective systems: (a)** $I_{He}^{T}$**, (b)** $I_{Be}^{T}$**, (c)** $I_{F}^{T}$**. The impurity atom has an approximate nominal valence state of -0, -4, and -1 for He, Be, and F, respectively. The electron configuration for** $I_{Be}^{T}$ **is** $1s^22s^22p^4$**.**

We can see that the Bader charge variation of matrix Li atoms is almost independent of the impurity types, even though these impurity atoms have different electronegativity and can lead to large lattice distortion. It is interesting that the Bader charge of the matrix Li increase slightly with the increasing pressure. This may be due to the distortion of the lattice induced by the interstitial impurity atoms, because it increases faster than in perfect FCC-Li (Table S7).

However, different impurity atoms have different influence on the ISQs. For system $I_{He}^{T}$, the impurity atom mainly leads to the increase of Bader charge in ISQ$^O$(NN) and ISQ$^T$(NNN) sites. For system $I_{Be}^{T}$, the impurity atom leads to the increase of Bader charge of impurity Be and ISQ$^O$(4NN) and the decrease of ISQ$^O$(NN). Whereas for system $I_{F}^{T}$, the impurity atom induce an increase of Bader charge of impurity F and ISQ$^O$(NNN) and a decrease of ISQ$^O$(NN).

**Table S8.** Averaged Bader charge ($q$) and its standard deviation ($\delta q$) among the same coordinate shell, and the difference with respect to the perfect FCC-Li ($\Delta Q$) at the same pressure for the matrix Li atoms, ISQ$^O$, and ISQ$^T$ sites surrounding the $I_{\text{He}}^T$ defect in a 3×3×3 FCC cubic supercell at the given pressures.

| site | 10 GPa | | | 20 GPa | | | 30 GPa | | | 40 GPa | | |
|---|---|---|---|---|---|---|---|---|---|---|---|---|
| | Bader charge (e) | | | Bader charge (e) | | | Bader charge (e) | | | Bader charge (e) | | |
| | $q$ | $\delta q$ | $\Delta Q$ | $q$ | $\delta q$ | $\Delta Q$ | $q$ | $\delta q$ | $\Delta Q$ | $q$ | $\delta q$ | $\Delta Q$ |
| Li(NN) | -0.77 | 0.00 | -0.01 | -0.75 | 0.00 | 0.02 | -0.72 | 0.00 | 0.04 | -0.70 | 0.00 | 0.06 |
| Li(NNN) | -0.76 | 0.00 | 0.00 | -0.74 | 0.00 | 0.03 | -0.71 | 0.00 | 0.05 | -0.70 | 0.00 | 0.07 |
| Li(3NN) | -0.77 | 0.00 | 0.00 | -0.74 | 0.00 | 0.03 | -0.71 | 0.00 | 0.05 | -0.70 | 0.00 | 0.07 |
| Li(4NN) | -0.77 | 0.00 | 0.00 | -0.74 | 0.00 | 0.03 | -0.71 | 0.00 | 0.05 | -0.70 | 0.00 | 0.07 |
| Li(M) | -0.76 | 0.00 | 0.00 | -0.73 | 0.00 | 0.03 | -0.71 | 0.00 | 0.05 | -0.70 | 0.00 | 0.07 |
| ISQ$^O$(NN) | 0.15 | 0.00 | -0.49 | 0.37 | 0.00 | -0.27 | 0.42 | 0.00 | -0.22 | 0.43 | 0.00 | -0.21 |
| ISQ$^O$(NNN) | 0.72 | 0.00 | 0.08 | 0.68 | 0.00 | 0.04 | 0.65 | 0.00 | 0.01 | 0.66 | 0.00 | 0.02 |
| ISQ$^O$(3NN) | 0.60 | 0.00 | -0.05 | 0.58 | 0.00 | -0.06 | 0.57 | 0.00 | -0.08 | 0.63 | 0.00 | -0.01 |
| ISQ$^O$(4NN) | 0.64 | 0.00 | -0.01 | 0.60 | 0.00 | -0.04 | 0.58 | 0.00 | -0.06 | 0.61 | 0.00 | -0.03 |
| ISQ$^O$(M) | 0.65 | 0.01 | 0.01 | 0.62 | 0.01 | -0.03 | 0.62 | 0.03 | -0.02 | 0.70 | 0.03 | 0.05 |
| ISQ$^T$(NN) | 0.00 | 0.00 | -0.06 | 0.00 | 0.00 | -0.06 | 0.00 | 0.00 | -0.06 | 0.00 | 0.00 | -0.06 |
| ISQ$^T$(NNN) | 0.19 | 0.00 | 0.13 | 0.13 | 0.06 | 0.07 | 0.11 | 0.08 | 0.05 | 0.10 | 0.10 | 0.04 |
| ISQ$^T$(3NN) | 0.04 | 0.01 | -0.02 | 0.03 | 0.03 | -0.03 | 0.02 | 0.02 | -0.04 | 0.00 | 0.00 | -0.06 |
| ISQ$^T$(4NN) | 0.04 | 0.00 | -0.03 | 0.04 | 0.00 | -0.02 | 0.03 | 0.01 | -0.03 | 0.00 | 0.00 | -0.06 |
| ISQ$^T$(5NN) | 0.06 | 0.00 | -0.00 | 0.06 | 0.00 | -0.00 | 0.00 | 0.00 | -0.06 | 0.00 | 0.00 | -0.06 |
| ISQ$^T$(6NN) | 0.07 | 0.01 | 0.01 | 0.07 | 0.01 | 0.01 | 0.06 | 0.00 | 0.00 | 0.00 | 0.00 | -0.06 |
| ISQ$^T$(7NN) | 0.06 | 0.00 | -0.00 | 0.06 | 0.00 | -0.00 | 0.06 | 0.00 | -0.00 | 0.01 | 0.00 | -0.05 |
| ISQ$^T$(M) | 0.06 | 0.00 | -0.00 | 0.06 | 0.00 | 0.00 | 0.06 | 0.01 | -0.00 | 0.01 | 0.01 | -0.05 |

**Table S9.** Averaged Bader charge ($q$) and its standard deviation ($\delta q$) among the same coordinate shell, and the difference with respect to the perfect FCC-Li ($\Delta Q$) at the same

pressure for the matrix Li atoms, ISQ$^O$, and ISQ$^T$ sites surrounding the $I_{Be}^T$ defect in a 3×3×3 FCC cubic supercell at the given pressures.

| site | 10 GPa Bader charge (e) | | | 20 GPa Bader charge (e) | | | 30 GPa Bader charge (e) | | | 40 GPa Bader charge (e) | | |
|---|---|---|---|---|---|---|---|---|---|---|---|---|
| | $q$ | $\delta q$ | $\Delta Q$ | $q$ | $\delta q$ | $\Delta Q$ | $q$ | $\delta q$ | $\Delta Q$ | $q$ | $\delta q$ | $\Delta Q$ |
| Li(NN) | -0.75 | 0.00 | 0.02 | -0.72 | 0.00 | 0.04 | -0.70 | 0.00 | 0.06 | -0.69 | 0.00 | 0.08 |
| Li(NNN) | -0.77 | 0.00 | -0.00 | -0.74 | 0.00 | 0.02 | -0.72 | 0.00 | 0.04 | -0.71 | 0.00 | 0.06 |
| Li(3NN) | -0.76 | 0.00 | 0.01 | -0.73 | 0.00 | 0.04 | -0.71 | 0.00 | 0.06 | -0.69 | 0.00 | 0.08 |
| Li(4NN) | -0.77 | 0.000 | 0.00 | -0.74 | 0.00 | 0.03 | -0.71 | 0.00 | 0.05 | -0.70 | 0.00 | 0.07 |
| Li(M) | -0.76 | 0.00 | 0.00 | -0.74 | 0.00 | 0.03 | -0.71 | 0.00 | 0.05 | -0.70 | 0.00 | 0.07 |
| ISQ$^O$(NN) | 0.00 | 0.00 | -0.64 | 0.00 | 0.00 | -0.64 | 0.00 | 0.00 | -0.64 | 0.00 | 0.00 | -0.64 |
| ISQ$^O$(NNN) | 0.65 | 0.00 | 0.07 | 0.61 | 0.00 | -0.03 | 0.57 | 0.00 | -0.07 | 0.54 | 0.00 | -0.10 |
| ISQ$^O$(3NN) | 0.67 | 0.00 | 0.03 | 0.72 | 0.00 | 0.08 | 0.85 | 0.00 | 0.20 | 0.86 | 0.00 | 0.21 |
| ISQ$^O$(4NN) | 0.77 | 0.00 | 0.13 | 0.77 | 0.00 | 0.13 | 0.79 | 0.00 | 0.14 | 0.83 | 0.00 | 0.18 |
| ISQ$^O$(M) | 0.63 | 0.03 | -0.01 | 0.63 | 0.05 | -0.02 | 0.64 | 0.06 | -0.00 | 0.66 | 0.08 | 0.02 |
| ISQ$^T$(NN) | 0.00 | 0.00 | -0.06 | 0.00 | 0.00 | -0.06 | 0.00 | 0.00 | -0.06 | 0.00 | 0.00 | -0.06 |
| ISQ$^T$(NNN) | 0.04 | 0.00 | -0.02 | 0.00 | 0.00 | -0.06 | 0.00 | 0.00 | -0.06 | 0.00 | 0.00 | -0.06 |
| ISQ$^T$(3NN) | 0.01 | 0.01 | -0.05 | 0.00 | 0.00 | -0.06 | 0.00 | 0.00 | -0.06 | 0.00 | 0.00 | -0.06 |
| ISQ$^T$(4NN) | 0.05 | 0.00 | -0.01 | 0.05 | 0.00 | -0.01 | 0.02 | 0.00 | -0.04 | 0.00 | 0.00 | -0.06 |
| ISQ$^T$(5NN) | 0.07 | 0.00 | 0.00 | 0.06 | 0.00 | -0.00 | 0.00 | 0.00 | -0.06 | 0.00 | 0.00 | -0.06 |
| ISQ$^T$(6NN) | 0.02 | 0.02 | -0.04 | 0.02 | 0.02 | -0.04 | 0.01 | 0.01 | -0.05 | 0.00 | 0.00 | -0.06 |
| ISQ$^T$(7NN) | 0.08 | 0.00 | 0.02 | 0.00 | 0.00 | -0.06 | 0.00 | 0.00 | -0.06 | 0.00 | 0.00 | -0.06 |
| ISQ$^T$(M) | 0.06 | 0.01 | 0.00 | 0.05 | 0.02 | -0.01 | 0.034 | 0.03 | -0.03 | 0.01 | 0.02 | -0.05 |

**Table S10.** Averaged Bader charge (*q*) and its standard deviation (δ*q*) among the same coordinate shell, and the difference with respect to the perfect FCC-Li (Δ*Q*) at the same pressure for the matrix Li atoms, ISQ$^O$, and ISQ$^T$ sites surrounding the $I_F^T$ defect in a 3×3×3 FCC cubic supercell at the given pressures.

| site | 10 GPa Bader charge (e) | | | 20 GPa Bader charge (e) | | | 30 GPa Bader charge (e) | | | 40 GPa Bader charge (e) | | |
|---|---|---|---|---|---|---|---|---|---|---|---|---|
| | $q$ | $\delta q$ | $\Delta Q$ | $q$ | $\delta q$ | $\Delta Q$ | $q$ | $\delta q$ | $\Delta Q$ | $q$ | $\delta q$ | $\Delta Q$ |
| Li(NN) | -0.80 | 0.00 | -0.03 | -0.78 | 0.00 | -0.01 | -0.76 | 0.00 | 0.01 | -0.74 | 0.00 | 0.02 |
| Li(NNN) | -0.76 | 0.00 | 0.00 | -0.74 | 0.00 | 0.03 | -0.72 | 0.00 | 0.05 | -0.70 | 0.00 | 0.07 |
| Li(3NN) | -0.76 | 0.00 | 0.00 | -0.74 | 0.00 | 0.03 | -0.71 | 0.00 | 0.05 | -0.69 | 0.000 | 0.07 |
| Li(4NN) | -0.76 | 0.00 | 0.00 | -0.74 | 0.00 | 0.03 | -0.71 | 0.00 | 0.05 | -0.70 | 0.00 | 0.07 |
| Li(M) | -0.76 | 0.00 | 0.00 | -0.74 | 0.00 | 0.03 | -0.71 | 0.00 | 0.05 | -0.70 | 0.00 | 0.07 |
| ISQ$^O$(NN) | 0.00 | 0.00 | -0.64 | 0.00 | 0.00 | -0.64 | 0.00 | 0.00 | -0.64 | 0.00 | 0.00 | -0.64 |
| ISQ$^O$(NNN) | 0.79 | 0.00 | 0.15 | 0.87 | 0.00 | 0.22 | 0.92 | 0.01 | 0.28 | 0.89 | 0.01 | 0.25 |
| ISQ$^O$(3NN) | 0.70 | 0.00 | 0.05 | 0.68 | 0.00 | 0.03 | 0.71 | 0.00 | 0.06 | 0.76 | 0.00 | 0.11 |
| ISQ$^O$(4NN) | 0.60 | 0.00 | -0.05 | 0.58 | 0.00 | -0.06 | 0.62 | 0.00 | -0.02 | 0.63 | 0.00 | -0.02 |
| ISQ$^O$(M) | 0.64 | 0.02 | 0.00 | 0.62 | 0.03 | -0.00 | 0.61 | 0.04 | -0.04 | 0.67 | 0.08 | 0.03 |
| ISQ$^T$(NN) | 0.00 | 0.00 | -0.06 | 0.00 | 0.00 | -0.06 | 0.00 | 0.00 | -0.06 | 0.00 | 0.00 | -0.06 |
| ISQ$^T$(NNN) | 0.03 | 0.00 | -0.03 | 0.04 | 0.00 | -0.02 | 0.00 | 0.00 | -0.06 | 0.00 | 0.00 | -0.06 |
| ISQ$^T$(3NN) | 0.08 | 0.03 | 0.02 | 0.08 | 0.04 | 0.02 | 0.07 | 0.04 | 0.01 | 0.00 | 0.00 | -0.06 |
| ISQ$^T$(4NN) | 0.06 | 0.00 | 0.00 | 0.05 | 0.00 | -0.01 | 0.01 | 0.01 | -0.05 | 0.01 | 0.01 | -0.05 |
| ISQ$^T$(5NN) | 0.06 | 0.00 | -0.01 | 0.00 | 0.00 | -0.06 | 0.00 | 0.00 | -0.06 | 0.00 | 0.00 | -0.06 |
| ISQ$^T$(6NN) | 0.06 | 0.00 | -0.00 | 0.03 | 0.03 | -0.03 | 0.00 | 0.00 | -0.06 | 0.00 | 0.00 | -0.06 |
| ISQ$^T$(7NN) | 0.06 | 0.00 | -0.00 | 0.06 | 0.00 | 0.00 | 0.00 | 0.00 | -0.06 | 0.00 | 0.00 | -0.06 |
| ISQ$^T$(M) | 0.06 | 0.01 | 0.00 | 0.06 | 0.00 | 0.00 | 0.06 | 0.01 | -0.00 | 0.00 | 0.01 | -0.06 |

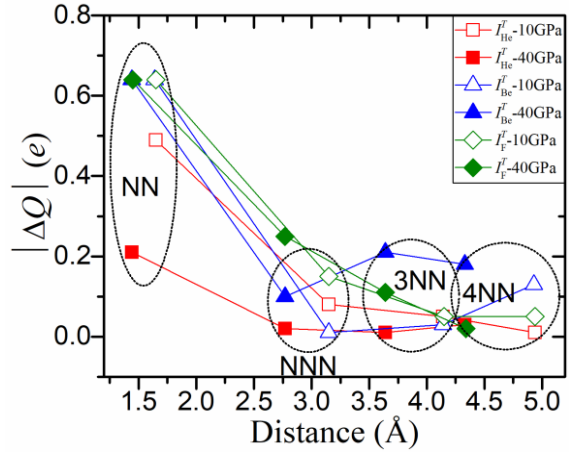

**Figure S12.** Attenuation of Bader charge of $O$ sites with the distance to the tetrahedral interstitial point defect.

One should note that the Bader charge of ISQ$^T$(NNN) in system $I^T_{He}$ has an obvious increase. This is related to the strong interplay between anionic ISQ$^O$(NN) and ISQ$^T$(NNN), because there is about 0.2-0.4 electrons locating at the ISQ$^O$(NN) for $I^T_{He}$, whereas it is zero for $I^T_{Be}$ and $I^T_F$.

## S4.2. Octahedral interstitial point defect: He, Be, and F

The detailed Bader charge variation of the matrix Li atoms, impurity atom, anionic $ISQ^O$ and $ISQ^T$ of the defective systems $I_{He}^O$, $I_{Be}^O$, and $I_F^O$ are shown in Fig. S13. The detailed values are listed in Tables S11-S13.

The octahedral interstitial impurity atoms have the similar influence on the Bader charge variation as the cases of occupying the $T$ sites when we compare the Bader charge of matrix Li atoms and distant neighboring ISQs. However, the influence on the ISQs is different from that of tetrahedral interstitial point defect. For system $I_{He}^O$, the impurity atom mainly leads to the increase of Bader charge in $ISQ^O(NN)$ and $ISQ^O(NNN)$ sites. For system $I_{Be}^O$, the impurity atom leads to the increase of Bader charge of impurity Be and $ISQ^O(NNN)$ and the decrease of $ISQ^O(NN)$. Whereas for system $I_F^O$, the impurity atom induces an increase of Bader charge of impurity F and a decrease of $ISQ^O(NNN)$.

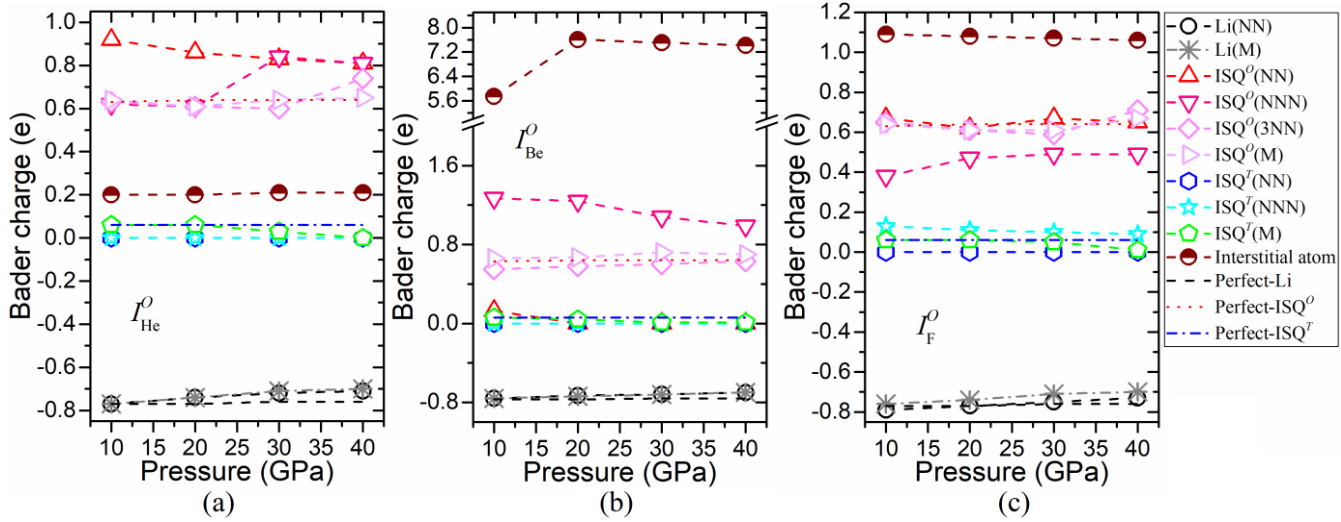

**Figure S13.** Variation of the Bader charge of the matrix Li atoms, impurity atom, anionic $ISQ^O$ and $ISQ^T$ of the defective systems: (a) $I_{He}^O$, (b) $I_{Be}^O$, (c) $I_F^O$. The impurity atom has an approximate nominal valence state of -0, -8, and -1 for He, Be, and F, respectively. The electron configuration for $I_{Be}^O$ is $1s^2 2s^2 2p^6 3s^2$ and violates the octet rule, whereas for $I_F^O$ is $1s^2 2s^2 2p^6$.

**Table S11.** Averaged Bader charge ($q$) and its standard deviation ($\delta q$) among the same coordinate shell, and the difference with respect to the perfect FCC-Li ($\Delta Q$) at the same pressure for the matrix Li atoms, $ISQ^O$, and $ISQ^T$ sites surrounding the $I_{He}^O$ defect in a 3×3×3 FCC cubic supercell at the given pressures.

| site | 10 GPa | | | 20 GPa | | | 30 GPa | | | 40 GPa | | |
| --- | --- | --- | --- | --- | --- | --- | --- | --- | --- | --- | --- | --- |
| | Bader charge (e) | | | Bader charge (e) | | | Bader charge (e) | | | Bader charge (e) | | |
| | $q$ | $\delta q$ | $\Delta Q$ | $q$ | $\delta q$ | $\Delta Q$ | $q$ | $\delta q$ | $\Delta Q$ | $q$ | $\delta q$ | $\Delta Q$ |
| Li(NN) | -0.77 | 0.00 | -0.01 | -0.74 | 0.00 | 0.02 | -0.72 | 0.00 | 0.04 | -0.71 | 0.00 | 0.06 |
| Li(NNN) | -0.76 | 0.00 | 0.00 | -0.73 | 0.00 | 0.03 | -0.71 | 0.00 | 0.05 | -0.69 | 0.00 | 0.07 |
| Li(3NN) | -0.77 | 0.00 | 0.00 | -0.74 | 0.00 | 0.03 | -0.71 | 0.00 | 0.05 | -0.70 | 0.00 | 0.07 |
| Li(4NN) | -0.77 | 0.00 | 0.00 | -0.73 | 0.00 | 0.03 | -0.71 | 0.00 | 0.05 | -0.69 | 0.00 | 0.07 |
| Li(M) | -0.77 | 0.00 | 0.00 | -0.74 | 0.00 | 0.03 | -0.71 | 0.00 | 0.05 | -0.70 | 0.00 | 0.07 |
| ISQ$^O$(NN) | 0.92 | 0.09 | 0.27 | 0.86 | 0.01 | 0.22 | 0.83 | 0.03 | 0.12 | 0.81 | 0.02 | 0.16 |
| ISQ$^O$(NNN) | 0.62 | 0.00 | -0.03 | 0.61 | 0.00 | -0.03 | 0.84 | 0.00 | 0.20 | 0.81 | 0.00 | 0.16 |
| ISQ$^O$(3NN) | 0.63 | 0.00 | -0.01 | 0.61 | 0.00 | -0.03 | 0.60 | 0.00 | -0.04 | 0.74 | 0.00 | 0.10 |
| ISQ$^O$(M) | 0.64 | 0.02 | -0.01 | 0.61 | 0.02 | -0.04 | 0.63 | 0.09 | -0.01 | 0.65 | 0.12 | 0.01 |
| ISQ$^T$(NN) | 0.00 | 0.00 | -0.06 | 0.00 | 0.00 | -0.06 | 0.00 | 0.00 | -0.06 | 0.00 | 0.00 | -0.06 |
| ISQ$^T$(NNN) | 0.00 | 0.00 | -0.06 | 0.00 | 0.00 | -0.06 | 0.00 | 0.00 | -0.06 | 0.00 | 0.00 | -0.06 |
| ISQ$^T$(M) | 0.06 | 0.01 | -0.00 | 0.06 | 0.00 | -0.00 | 0.03 | 0.03 | -0.03 | 0.00 | 0.00 | -0.06 |

**Table S12.** Averaged Bader charge ($q$) and its standard deviation ($\delta q$) among the same coordinate shell, and the difference with respect to the perfect FCC-Li ($\Delta Q$) at the same pressure for the matrix Li atoms, ISQ$^O$, and ISQ$^T$ sites surrounding the $I_{Be}^O$ defect in a 3×3×3 FCC cubic supercell at the given pressures.

| site | 10 GPa | | | 20 GPa | | | 30 GPa | | | 40 GPa | | |
| --- | --- | --- | --- | --- | --- | --- | --- | --- | --- | --- | --- | --- |
| | Bader charge (e) | | | Bader charge (e) | | | Bader charge (e) | | | Bader charge (e) | | |
| | $q$ | $\delta q$ | $\Delta Q$ | $q$ | $\delta q$ | $\Delta Q$ | $q$ | $\delta q$ | $\Delta Q$ | $q$ | $\delta q$ | $\Delta Q$ |
| Li(NN) | -0.76 | 0.00 | 0.01 | -0.73 | 0.00 | 0.03 | -0.72 | 0.00 | 0.05 | -0.70 | 0.00 | 0.07 |
| Li(NNN) | -0.77 | 0.00 | -0.01 | -0.75 | 0.00 | 0.02 | -0.73 | 0.00 | 0.04 | -0.71 | 0.00 | 0.05 |
| Li(3NN) | -0.76 | 0.00 | 0.00 | -0.73 | 0.00 | 0.03 | -0.71 | 0.00 | 0.05 | -0.69 | 0.00 | 0.07 |
| Li(4NN) | -0.76 | 0.00 | 0.00 | -0.74 | 0.00 | 0.03 | -0.71 | 0.00 | 0.05 | -0.70 | 0.00 | 0.07 |
| Li(M) | -0.76 | 0.00 | 0.00 | -0.74 | 0.00 | 0.03 | -0.71 | 0.00 | 0.05 | -0.70 | 0.00 | 0.07 |
| ISQ$^O$(NN) | 0.13 | 0.00 | -0.52 | 0.00 | 0.00 | -0.64 | 0.00 | 0.00 | -0.64 | 0.00 | 0.00 | -0.64 |
| ISQ$^O$(NNN) | 1.27 | 0.11 | 0.63 | 1.24 | 0.00 | 0.60 | 1.08 | 0.00 | 0.43 | 0.99 | 0.00 | 0.35 |
| ISQ$^O$(3NN) | 0.55 | 0.00 | -0.10 | 0.58 | 0.00 | -0.06 | 0.60 | 0.00 | -0.04 | 0.63 | 0.00 | -0.01 |
| ISQ$^O$(M) | 0.66 | 0.05 | 0.02 | 0.67 | 0.10 | 0.03 | 0.72 | 0.15 | 0.07 | 0.70 | 0.13 | 0.05 |
| ISQ$^T$(NN) | 0.00 | 0.00 | -0.06 | 0.00 | 0.00 | -0.06 | 0.00 | 0.00 | -0.06 | 0.00 | 0.00 | -0.06 |
| ISQ$^T$(NNN) | 0.00 | 0.00 | -0.06 | 0.00 | 0.00 | -0.06 | 0.00 | 0.00 | -0.06 | 0.00 | 0.00 | -0.06 |
| ISQ$^T$(M) | 0.06 | 0.02 | 0.00 | 0.04 | 0.03 | -0.02 | 0.01 | 0.03 | -0.05 | 0.01 | 0.02 | -0.05 |

**Table S13.** Averaged Bader charge ($q$) and its standard deviation ($\delta q$) among the same coordinate shell, and the difference with respect to the perfect FCC-Li ($\Delta Q$) at the same pressure for the matrix Li atoms, ISQ$^O$, and ISQ$^T$ sites surrounding the $I_F^O$ defect in a 3×3×3 FCC cubic supercell at the given pressures.

| site | 10 GPa | 20 GPa | 30 GPa | 40 GPa |
| --- | --- | --- | --- | --- |

|  | Bader charge (e) | | | Bader charge (e) | | | Bader charge (e) | | | Bader charge (e) | | |
|---|---|---|---|---|---|---|---|---|---|---|---|---|
|  | $q$ | $\delta q$ | $\Delta Q$ | $q$ | $\delta q$ | $\Delta Q$ | $q$ | $\delta q$ | $\Delta Q$ | $q$ | $\delta q$ | $\Delta Q$ |
| Li(NN) | -0.79 | 0.00 | -0.02 | -0.77 | 0.00 | -0.00 | -0.75 | 0.00 | 0.02 | -0.73 | 0.00 | 0.04 |
| Li(NNN) | -0.76 | 0.03 | 0.01 | -0.73 | 0.00 | 0.03 | -0.71 | 0.00 | 0.05 | -0.69 | 0.00 | 0.07 |
| Li(3NN) | -0.77 | 0.00 | 0.00 | -0.74 | 0.00 | 0.03 | -0.71 | 0.00 | 0.05 | -0.70 | 0.00 | 0.07 |
| Li(4NN) | -0.76 | 0.00 | 0.00 | -0.73 | 0.00 | 0.03 | -0.71 | 0.00 | 0.05 | -0.70 | 0.00 | 0.07 |
| Li(M) | -0.76 | 0.00 | 0.00 | -0.74 | 0.00 | 0.03 | -0.71 | 0.00 | 0.05 | -0.70 | 0.00 | 0.07 |
| ISQ$^O$(NN) | 0.67 | 0.00 | 0.02 | 0.62 | 0.00 | -0.02 | 0.67 | 0.04 | 0.03 | 0.65 | 0.00 | 0.01 |
| ISQ$^O$(NNN) | 0.38 | 0.00 | -0.27 | 0.47 | 0.00 | -0.18 | 0.49 | 0.00 | -0.16 | 0.49 | 0.00 | -0.16 |
| ISQ$^O$(3NN) | 0.65 | 0.00 | 0.00 | 0.61 | 0.00 | -0.03 | 0.59 | 0.00 | -0.05 | 0.71 | 0.00 | 0.06 |
| ISQ$^O$(M) | 0.64 | 0.02 | -0.00 | 0.61 | 0.02 | -0.03 | 0.61 | 0.03 | -0.03 | 0.67 | 0.08 | 0.02 |
| ISQ$^T$(NN) | 0.00 | 0.00 | -0.06 | 0.00 | 0.00 | -0.06 | 0.00 | 0.00 | -0.06 | 0.00 | 0.00 | -0.06 |
| ISQ$^T$(NNN) | 0.13 | 0.00 | 0.06 | 0.11 | 0.00 | 0.05 | 0.10 | 0.00 | 0.04 | 0.09 | 0.00 | 0.03 |
| ISQ$^T$(M) | 0.06 | 0.00 | -0.00 | 0.06 | 0.00 | -0.00 | 0.05 | 0.02 | -0.01 | 0.01 | 0.01 | -0.05 |

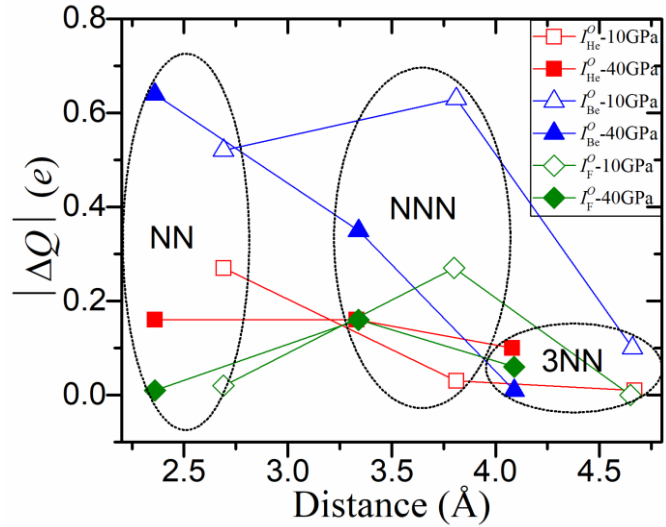

**Figure S14. Attenuation of Bader charge of $O$ sites with the distance to the octahedral interstitial point defect.**

Figure S14 shows the attenuation degree of Bader charge of $O$ sites with the distance to the point defect. These results agree well with the influence areas as shown in Fig. S6. Because $T$ site is not the favorable electron localization interstice, impurity atoms have a negligible influence to them.

Another should be noted phenomenon is that the Bader charge of ISQ$^O$(3NN) has a small increase at 40 GPa for systems $I_{He}^O$ and $I_F^O$. This should be attributed to the increase of the electronegativity[1] of He and F, as well as the increase of the influenced area by impurity atoms (see Table S1). For example, the electronegativity and the relative influenced area for system $I_{Be}^O$ increase from ~9 to ~10 eV/$e$ and from 36% to 39% with the pressure increased from 10 GPa to 40 GPa, respectively.

**S4.3. Self-interstitial point defect: Li**

The detailed Bader charge variation of the matrix Li atoms, impurity atom, anionic $ISQ^O$ and $ISQ^T$ of the self-interstitial defective systems $I_{Li}^T$ and $I_{Li}^O$ are shown in Fig. S15. The specific values are listed in Tables S14-S15.

We can see that the self-interstitial Li has similar Bader charge as the matrix Li atoms for system $I_{Li}^T$ and $I_{Li}^O$. However, the impurity Li atom occupying different interstitial sites has different influence on the ISQs. For system $I_{Li}^T$, the influence of tetrahedral self-interstitial Li atoms on the electron localization of distant neighboring Li/ISQs is similar to other impurity atoms, i.e. the increase of the Bader charge of nearby $ISQ^O$ and $ISQ^T$. For system $I_{Li}^O$, the Bader charges of $ISQ^O$(NN) and $ISQ^T$(NN) have a sharp increase and decrease at 20-30 GPa, respectively, and the Bader charge of $ISQ^O$(NNN) has an obvious decrease with the increasing pressure. Obviously, this is a consequence of the redistribution of the anionic electrons between ISQs driven by pressure.

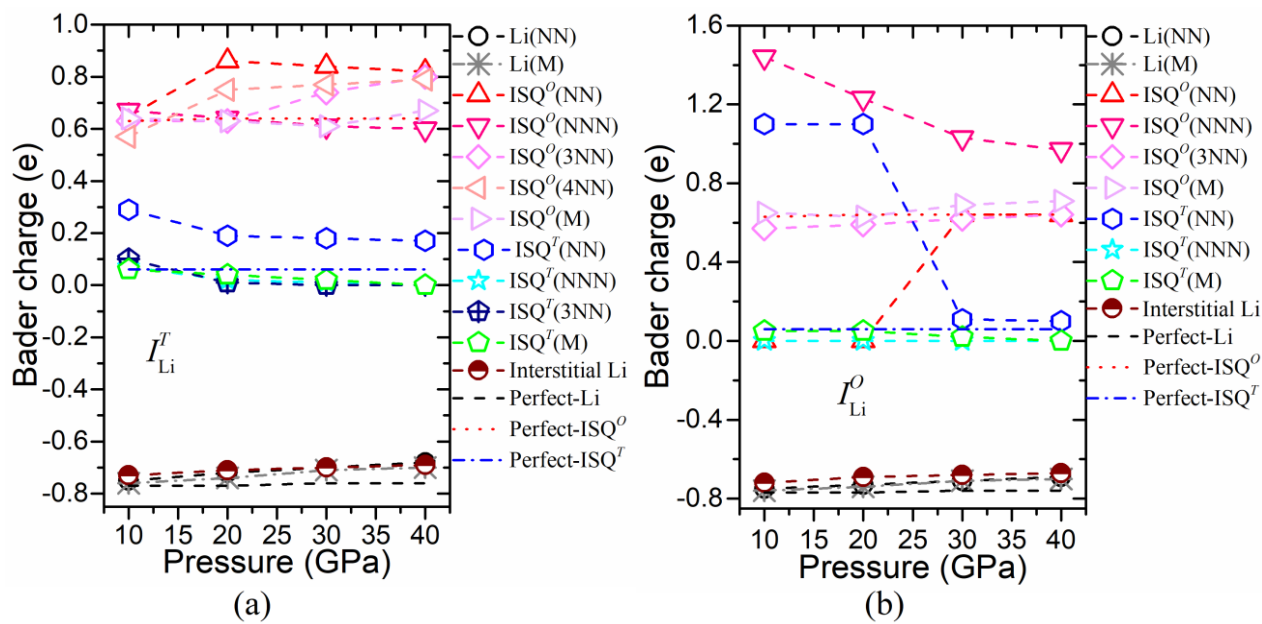

**Figure S15.** **Variation of the Bader charge of the matrix Li atoms, impurity atom, anionic $ISQ^O$ and $ISQ^T$ of the defective systems: (a)** $I_{Li}^T$**, (b)** $I_{Li}^O$**.**

**Table S14.** Averaged Bader charge (*q*) and its standard deviation (δ*q*) among the same coordinate shell, and the difference with respect to the perfect FCC-Li (Δ*Q*) at the same pressure for the matrix Li atoms, ISQ$^O$, and ISQ$^T$ sites surrounding the $I_{Li}^T$ defect in a 3×3×3 FCC cubic supercell at the given pressures.

| site | 10 GPa | | | 20 GPa | | | 30 GPa | | | 40 GPa | | |
|---|---|---|---|---|---|---|---|---|---|---|---|---|
| | Bader charge (e) | | | Bader charge (e) | | | Bader charge (e) | | | Bader charge (e) | | |
| | *q* | δ*q* | Δ*Q* | *q* | δ*q* | Δ*Q* | *q* | δ*q* | Δ*Q* | *q* | δ*q* | Δ*Q* |
| Li(NN) | -0.75 | 0.00 | 0.02 | -0.72 | 0.00 | 0.05 | -0.70 | 0.00 | 0.07 | -0.68 | 0.00 | 0.08 |
| Li(NNN) | -0.77 | 0.00 | -0.00 | -0.74 | 0.00 | 0.02 | -0.72 | 0.00 | 0.05 | -0.70 | 0.00 | 0.06 |
| Li(3NN) | -0.76 | 0.00 | 0.01 | -0.73 | 0.00 | 0.03 | -0.71 | 0.00 | 0.05 | -0.69 | 0.00 | 0.07 |
| Li(4NN) | -0.77 | 0.00 | -0.00 | -0.74 | 0.00 | 0.03 | -0.72 | 0.00 | 0.05 | -0.70 | 0.00 | 0.07 |
| Li(M) | -0.76 | 0.00 | 0.00 | -0.74 | 0.00 | 0.03 | -0.71 | 0.00 | 0.05 | -0.70 | 0.00 | 0.07 |
| ISQ$^O$(NN) | 0.65 | 0.00 | 0.01 | 0.86 | 0.00 | 0.22 | 0.84 | 0.00 | 0.20 | 0.82 | 0.00 | 0.18 |
| ISQ$^O$(NNN) | 0.67 | 0.00 | 0.03 | 0.64 | 0.00 | -0.01 | 0.61 | 0.00 | -0.03 | 0.60 | 0.00 | -0.05 |
| ISQ$^O$(3NN) | 0.63 | 0.02 | -0.02 | 0.63 | 0.00 | -0.01 | 0.74 | 0.00 | 0.10 | 0.80 | 0.00 | 0.16 |
| ISQ$^O$(4NN) | 0.57 | 0.00 | -0.07 | 0.75 | 0.00 | 0.10 | 0.77 | 0.00 | 0.13 | 0.79 | 0.00 | 0.14 |
| ISQ$^O$(M) | 0.64 | 0.02 | -0.01 | 0.63 | 0.05 | -0.01 | 0.61 | 0.05 | -0.03 | 0.67 | 0.09 | 0.03 |
| ISQ$^T$(NN) | 0.29 | 0.00 | 0.22 | 0.19 | 0.00 | 0.13 | 0.19 | 0.01 | 0.13 | 0.17 | 0.00 | 0.11 |
| ISQ$^T$(NNN) | 0.07 | 0.00 | 0.01 | 0.02 | 0.03 | -0.04 | 0.01 | 0.00 | -0.05 | 0.00 | 0.00 | -0.06 |
| ISQ$^T$(3NN) | 0.10 | 0.07 | 0.04 | 0.01 | 0.01 | -0.05 | 0.00 | 0.00 | -0.06 | 0.00 | 0.00 | -0.06 |
| ISQ$^T$(4NN) | 0.04 | 0.00 | -0.02 | 0.04 | 0.00 | -0.02 | 0.01 | 0.00 | -0.05 | 0.00 | 0.00 | -0.06 |
| ISQ$^T$(5NN) | 0.06 | 0.00 | -0.00 | 0.06 | 0.00 | -0.00 | 0.02 | 0.01 | -0.04 | 0.00 | 0.00 | -0.06 |
| ISQ$^T$(6NN) | 0.03 | 0.03 | -0.03 | 0.03 | 0.03 | -0.03 | 0.03 | 0.03 | -0.03 | 0.00 | 0.00 | -0.06 |
| ISQ$^T$(7NN) | 0.08 | 0.00 | 0.02 | 0.00 | 0.00 | -0.06 | 0.00 | 0.00 | -0.06 | 0.00 | 0.00 | -0.06 |
| ISQ$^T$(M) | 0.06 | 0.01 | 0.00 | 0.06 | 0.02 | -0.01 | 0.05 | 0.02 | -0.01 | 0.06 | 0.01 | -0.06 |

**Table S15.** Averaged Bader charge (*q*) and its standard deviation (δ*q*) among the same coordinate shell, and the difference with respect to the perfect FCC-Li (Δ*Q*) at the same pressure for the matrix Li atoms, ISQ$^O$, and ISQ$^T$ sites surrounding the $I_{Li}^O$ defect in a 3×3×3 FCC cubic supercell at the given pressures.

| site | 10 GPa | | | 20 GPa | | | 30 GPa | | | 40 GPa | | |
|---|---|---|---|---|---|---|---|---|---|---|---|---|
| | Bader charge (e) | | | Bader charge (e) | | | Bader charge (e) | | | Bader charge (e) | | |
| | *q* | δ*q* | Δ*Q* | *q* | δ*q* | Δ*Q* | *q* | δ*q* | Δ*Q* | *q* | δ*q* | Δ*Q* |
| Li(NN) | -0.75 | 0.00 | 0.01 | -0.73 | 0.00 | 0.04 | -0.71 | 0.00 | 0.06 | -0.69 | 0.00 | 0.08 |
| Li(NNN) | -0.77 | 0.00 | -0.01 | -0.74 | 0.00 | 0.02 | -0.72 | 0.00 | 0.04 | -0.70 | 0.00 | 0.06 |
| Li(3NN) | -0.76 | 0.00 | 0.00 | -0.73 | 0.00 | 0.03 | -0.71 | 0.00 | 0.05 | -0.69 | 0.00 | 0.07 |
| Li(4NN) | -0.76 | 0.00 | 0.00 | -0.74 | 0.00 | 0.03 | -0.71 | 0.00 | 0.05 | -0.70 | 0.00 | 0.07 |
| Li(M) | -0.76 | 0.00 | 0.00 | -0.74 | 0.00 | 0.03 | -0.71 | 0.00 | 0.05 | -0.70 | 0.00 | 0.07 |
| ISQ$^O$(NN) | 0.00 | 0.00 | -0.64 | 0.00 | 0.00 | -0.64 | 0.64 | 0.00 | -0.01 | 0.64 | 0.00 | -0.00 |
| ISQ$^O$(NNN) | 1.44 | 0.00 | 0.80 | 1.23 | 0.00 | 0.59 | 1.03 | 0.00 | 0.39 | 0.97 | 0.00 | 0.33 |
| ISQ$^O$(3NN) | 0.57 | 0.00 | -0.08 | 0.59 | 0.00 | -0.05 | 0.62 | 0.00 | -0.02 | 0.64 | 0.00 | -0.01 |
| ISQ$^O$(M) | 0.65 | 0.06 | 0.00 | 0.63 | 0.08 | -0.01 | 0.69 | 0.18 | 0.05 | 0.70 | 0.15 | 0.06 |
| ISQ$^T$(NN) | 1.10 | 0.44 | 1.04 | 1.10 | 0.40 | 1.04 | 0.11 | 0.00 | 0.05 | 0.10 | 0.00 | 0.04 |
| ISQ$^T$(NNN) | 0.00 | 0.00 | -0.06 | 0.00 | 0.00 | -0.06 | 0.00 | 0.00 | -0.06 | 0.00 | 0.00 | -0.06 |
| ISQ$^T$(M) | 0.05 | 0.02 | -0.01 | 0.05 | 0.03 | -0.02 | 0.02 | 0.03 | -0.04 | 0.00 | 0.01 | -0.06 |

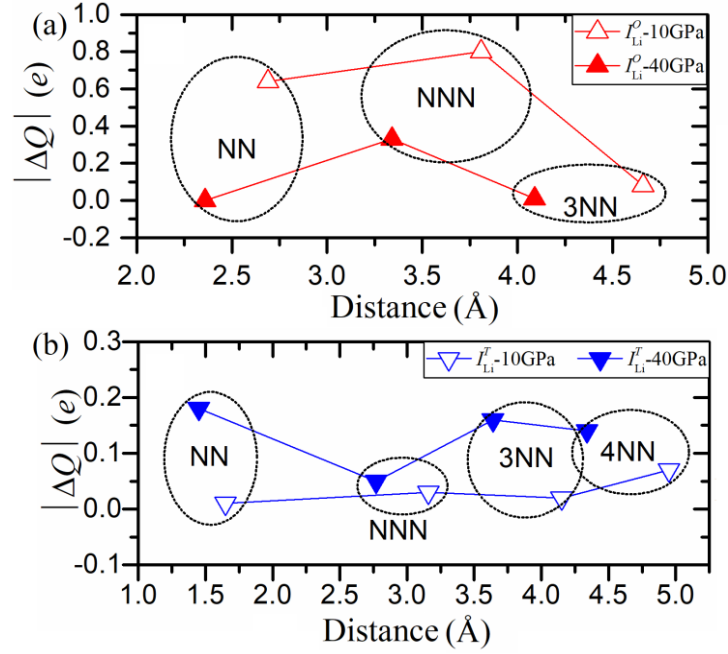

**Figure S16.** Attenuation of Bader charge of *O* sites with the distance to the octahedral/tetrahedral self-interstitial point defect.

We can make it clear from above discussion that (a) the radius of the influenced area by different interstitial impurity atoms are ~20%-40% of the lattice length of a 3×3×3 cubic supercell within the studied pressure range (10-40 GPa); (b) the variation of the valence state of matrix Li atoms is independent of the impurity atom type and their occupation site; (c) impurity atoms mainly have an influence on the Bader charge of nearby *O* and *T* sites, and the change in *T* sites is far less than that of the *O* sites; (d) the influence to the far neighboring *O* sites will increase slightly with increasing pressure. These rules are also applicable to other electrides.

## S5. Distance of Li-F in crystalline LiF and defective FCC-Li

Lithium fluoride (LiF) crystallizes in NaCl rocksalt structure at atmospheric pressure. Previous experiment and theory work[2][3] predicated that it undergoes no phase transformation between 0-100 GPa. The zero-pressure volume $V_0$ of LiF given in Ref.[2] and therein is 65.02-68.62 Å$^3$, our calculation gives a result of $V_0$=66.92 Å$^3$ at atmospheric pressure. We further

S21

calculated the Li-F bond length in crystalline LiF by using DFT-PBE method. The results are shown in Fig. S17, and are compared with the nearest neighboring Li-F distance in system $I_\text{F}^O$.

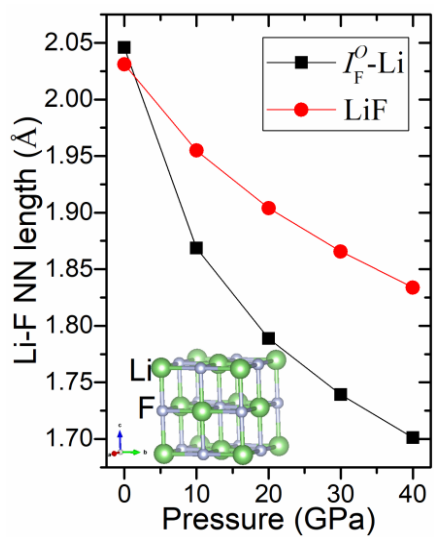

**Figure S17.** **Variation of the nearest neighboring Li-F distance in crystalline LiF and in defective FCC-Li with $I_\text{F}^O$ between 0-40 GPa, inset shows the crystalline structure of LiF.**

## S6. Difference between electrides and covalent compounds

Figure S18 shows the electron localization function (ELF) of different compounds. We can see that the unique feature of electrides is characterized by the unconventional localization of excess electrons away from the atomic nuclei, as shown in Fig. S18a. Different from conventional covalent compounds in which the valence electrons are shared by the bonding atoms (see Fig. S18b and Fig. S18c) and ionic compounds in which the valence electrons transferred from cations to anions (see Fig. S18d), the excess electrons in electrides are confined and localized into the lattice interstitial site.

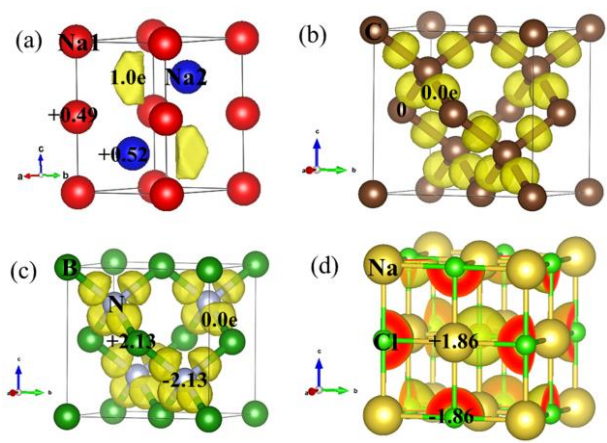

**Figure S18.** Electron localization function (ELF) of (a) electride *hP*4-Na at 200 GPa, (b) covalent bonds in diamond at 0 GPa, (c) covalent bonds in cubic boron nitride (cBN) at 0 GPa, (d) ionic NaCl at 0 GPa, all with isosurface=0.80. The charge state of atoms and the electron-localized sites are also indicated.

Obviously, ELF is not enough to describe and analysis the localization behavior of electrons in electrides since covalent bonds and ISQs are similar in ELF. The key difference between covalent bond and ISQ is whether there are actual electrons locating at the maximal center of ELF or not, as demonstrated clearly in Fig. S18. Namely, the ISQ must be negatively charged in electride. Ignoring this difference has led to confusion in literature where some covalent compounds were wrongly classed as electrides. For example, as shown in Fig. S19, although there are maximal center of ELF in $CeH_2$[4] and LaScSi[5], there are no actual electrons locating at these sites. Therefore, $CeH_2$ and LaScSi should not be regarded as electrides.

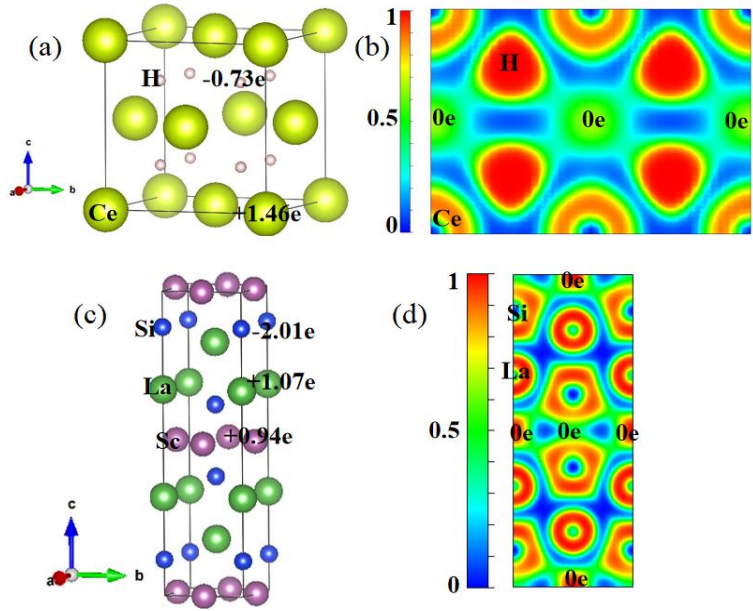

**Figure S19.** Crystal structure of (a) $CeH_2$ and (c) LaScSi. (b) and (d) are the ELF of the (1 1 0) plane in (a) and (c). The charge state of atoms and the electron-localized sites are also indicated.

### S7. ELF of defective FCC-Li in 3×3×3 cubic supercell

Figure S20 shows the ELF of the defective FCC-Li in a 3×3×3 cubic supercell, which is the supplement of Figs. 6-9 in the main text. Figure S21 shows the ELF of the (1 1 0) plane in these defective systems. We can see that the local ELF around the impurity Be atom has a visible 3-D character (Fig. S21b and S21e). Obviously, this is closely related to the electronegativity of impurity atoms. The local ELF around the impurity Li atom also shows a 3-D character (Fig. S21g and S21h), though it is weaker than Be.

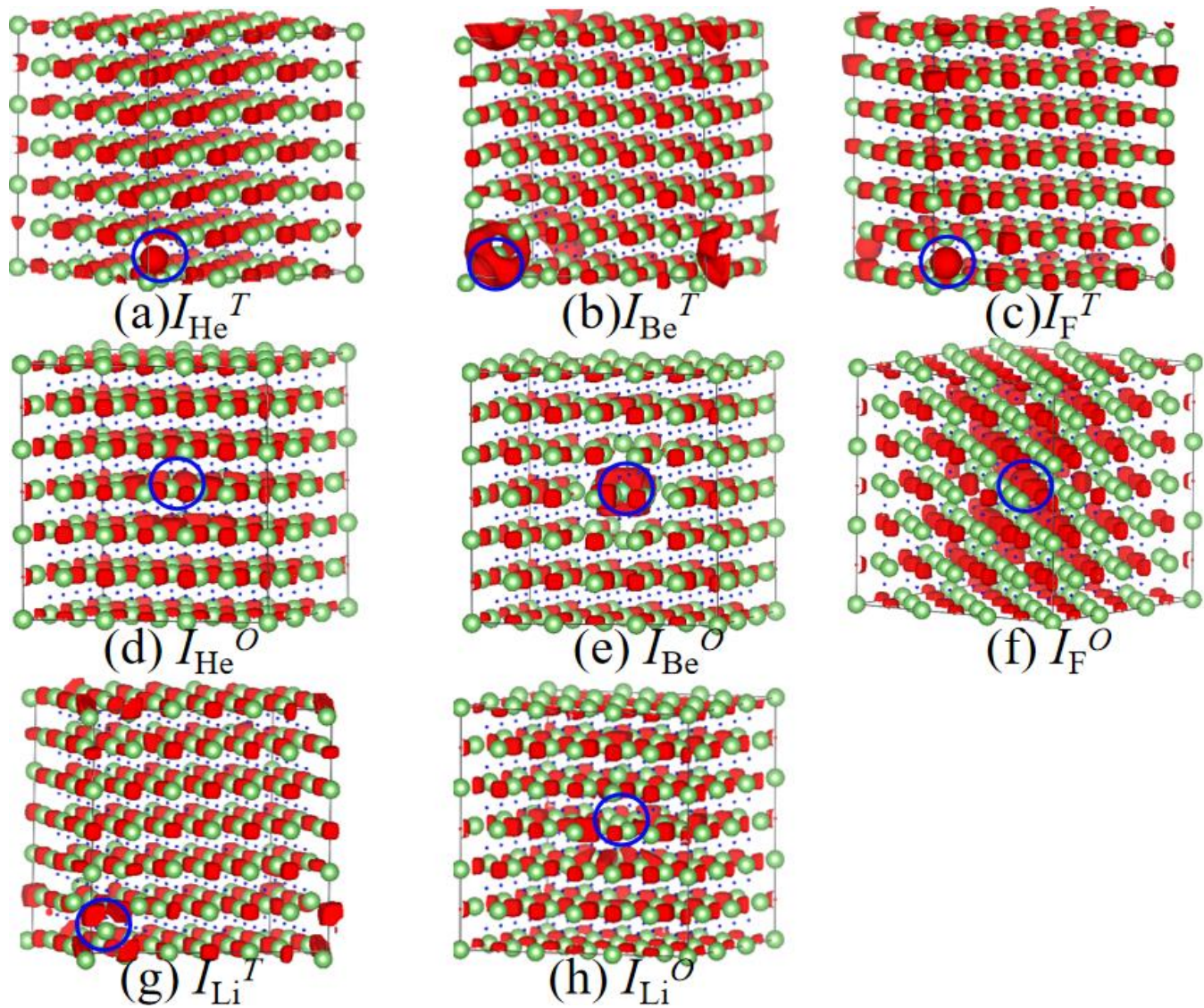

**Figure S20.** Calculated ELF (isosurface=0.75) of defective FCC-Li with different interstitial impurity in a cubic 3×3×3 FCC supercell at 10 GPa. The big green, small red and small blue spheres represent Li atom, $O$ and $T$ sites, respectively. The point defects are marked with a blue circle.

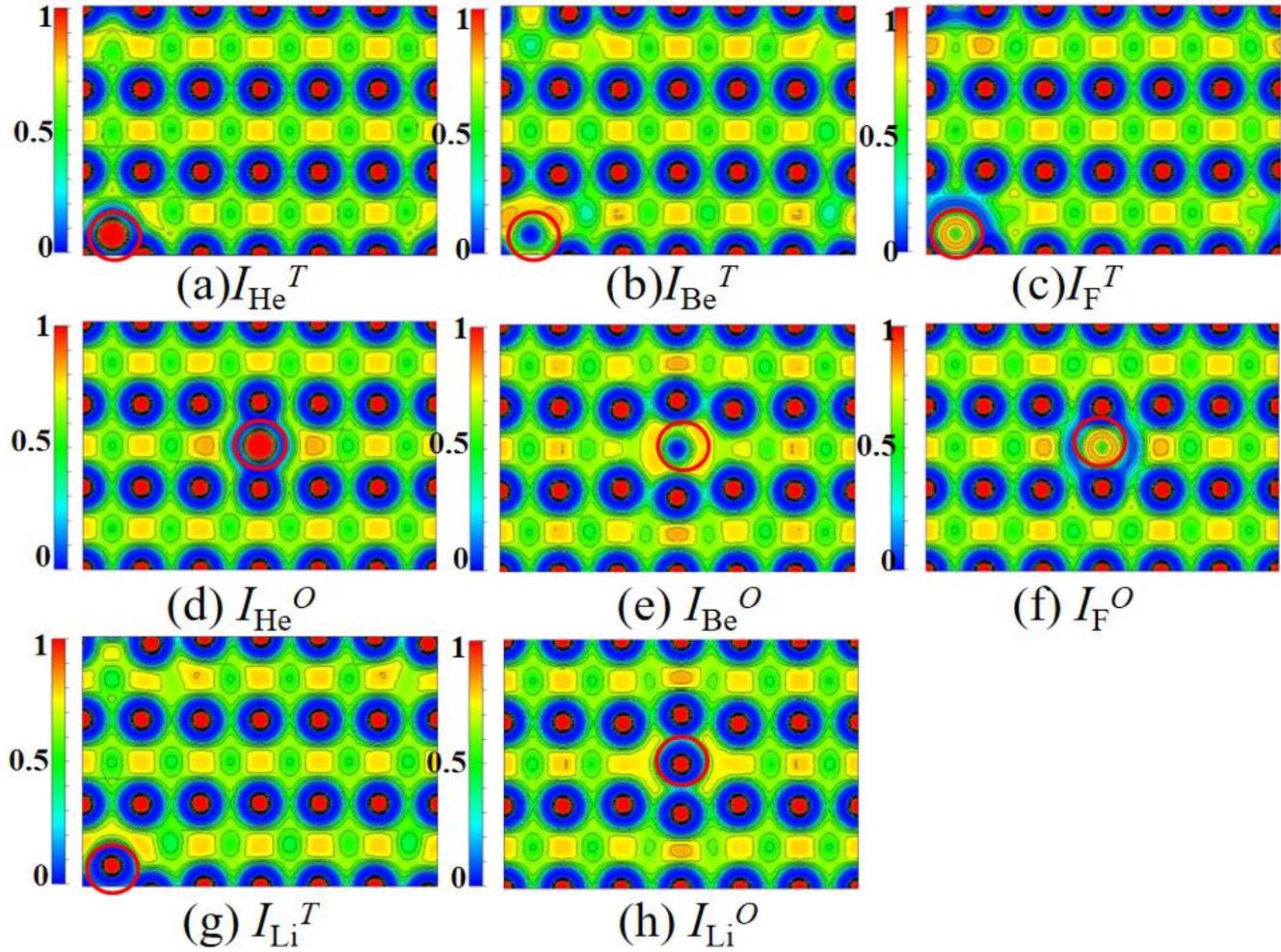

**Figure S21.** Calculated ELF in the (1 1 0) plane of defective FCC-Li as shown in Fig. S20. The point defects are marked with red circles.

### S8. DOS and band structure of defective FCC-Li in 3×3×3 cubic supercell

Figure S22 shows the total electronic density of state (DOS) of the defective FCC-Li in a 3×3×3 cubic supercell, which are compared with the perfect FCC-Li. We can see that the interstitial impurity atoms have a small contribution to the total DOS near the Fermi surface. The most obvious influence on the DOS is at around 4-5 eV. The local projected DOS, as shown in Fig. S23, support this conclusion.

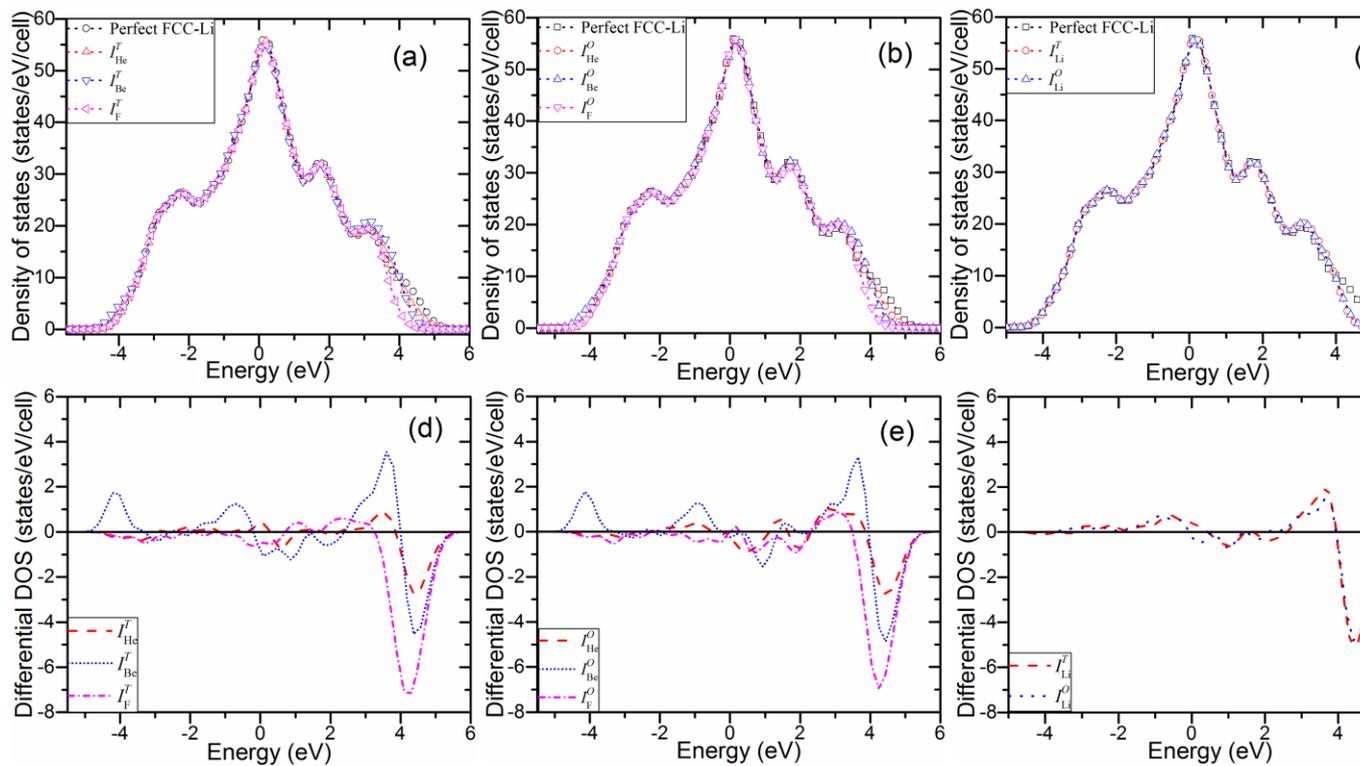

**Figure S22.** Total DOS of the defective FCC-Li compared with perfect FCC-Li at 10 GPa (a-c); The differential DOS between the defective FCC-Li and the perfect FCC-Li at 10 GPa (d-f).

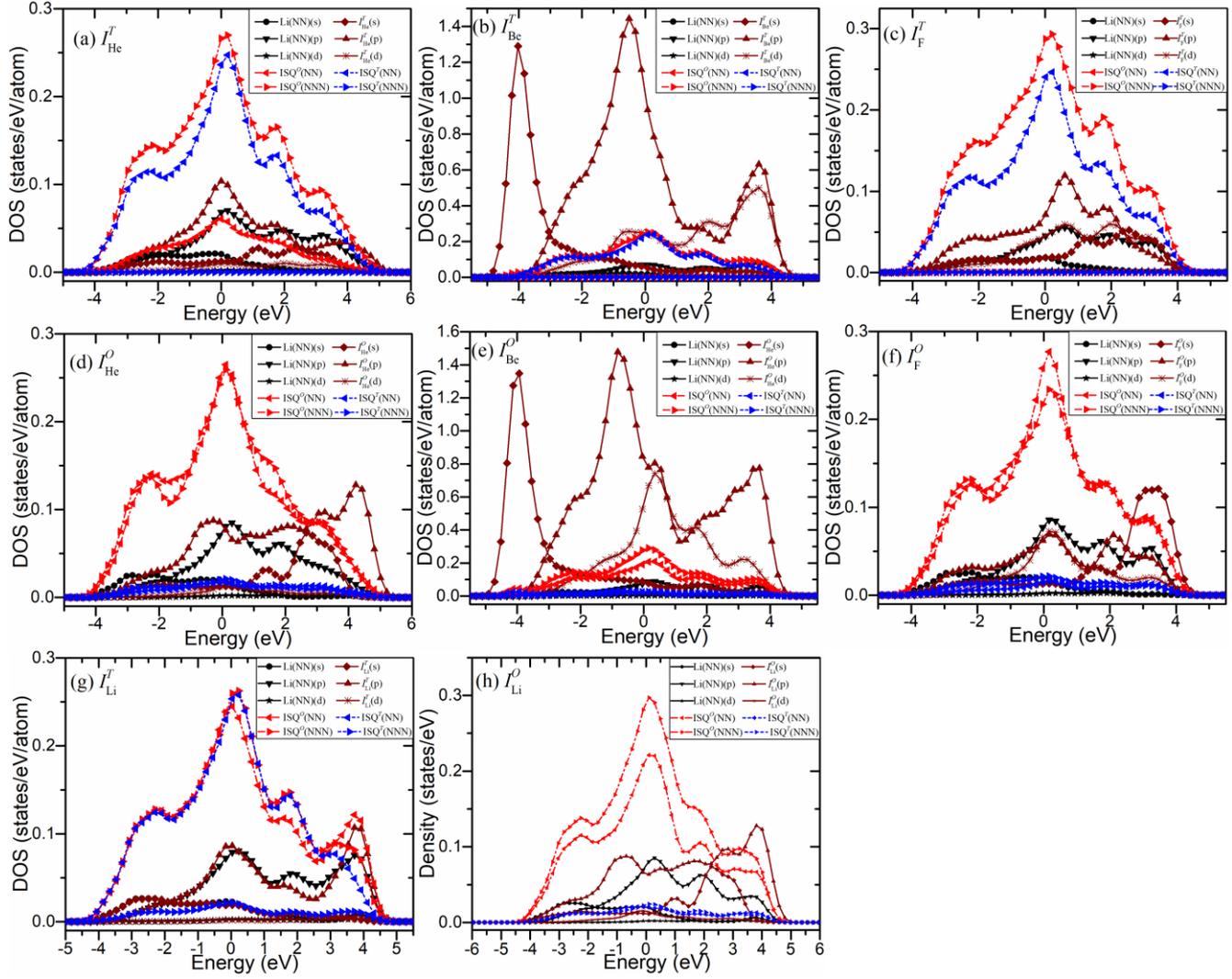

**Figure S23.** Nearest- and next-nearest-ISQs contribution to the DOS of the defective systems compared with the point-defects and matrix Li atoms at 10 GPa.

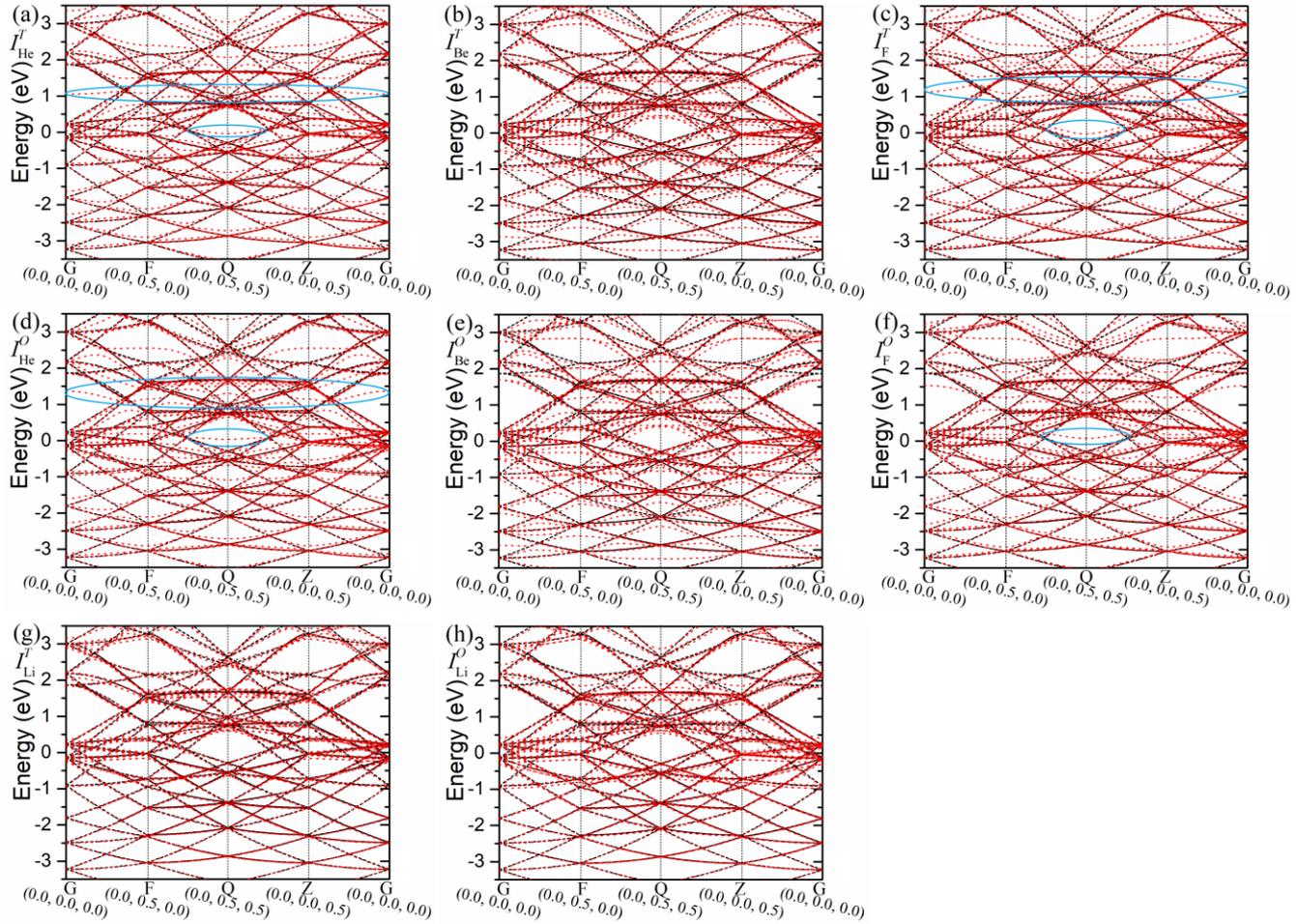

**Figure S24.** **Band structure of the defective systems compared with the perfect FCC-Li at 10 GPa. Red lines represent the defective systems, and black lines represent the perfect FCC-Li.**

Figure S24 shows the band structure of the defective systems compared with the perfect FCC-Li, which is the supplement of Fig. 10 in the main text. We can see that the impurity He and F have a main influence on the energy bands at near and above the Fermi surface, and they induce new bands near the Fermi surface and the Gamma point. The impurity Be influences the whole energy range, while interstitial Li impurity only has an obvious influence above the Fermi surface. Interestingly, these interstitial impurity atoms all enhance the electron localization characteristics in dense FCC-Li.

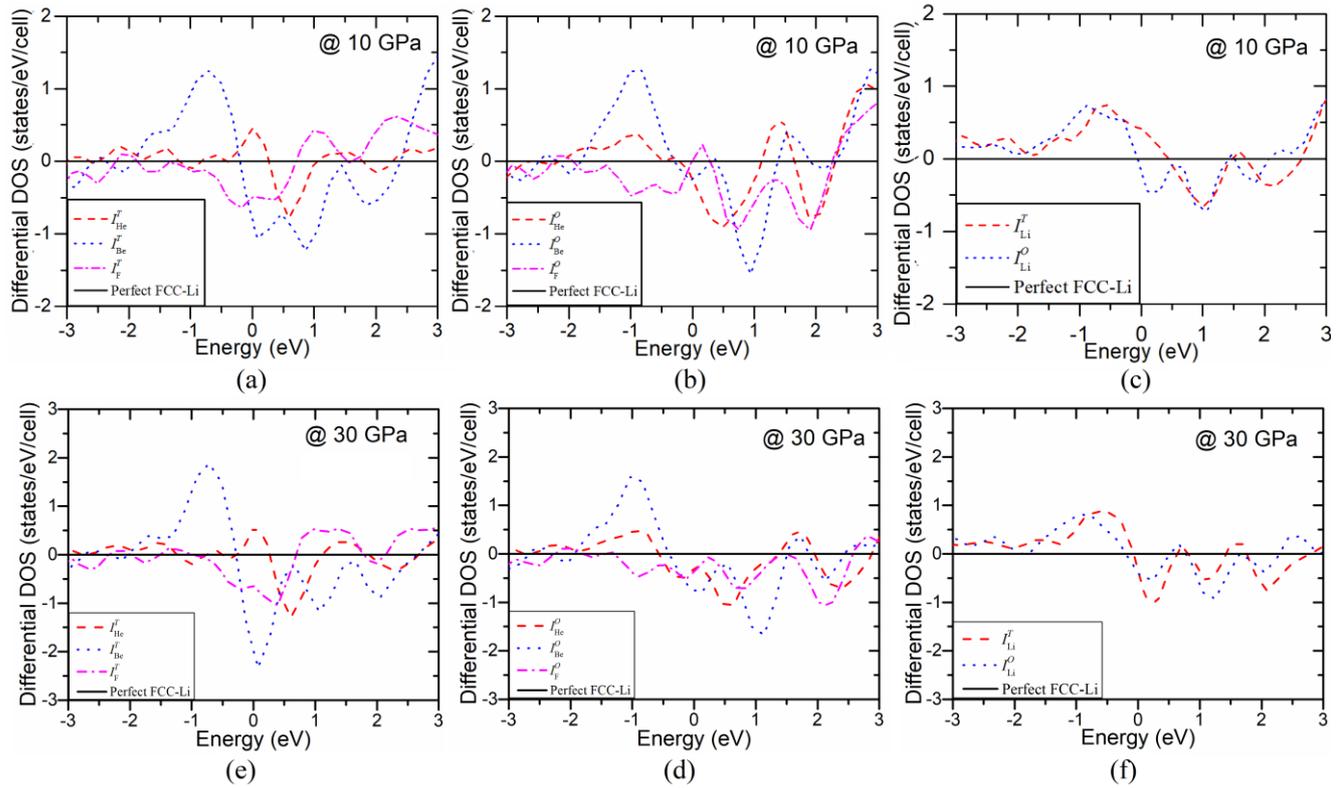

**Figure S25.** Differential DOS of the defective systems with respect to the perfect FCC-Li.

Figure S25 shows the comparison of the differential DOS of these defective systems at 10 GPa and 30 GPa. We can see that pressure has a small influence on the matrix electronic structure of these defective systems.

## S9. Results of high defects density

To have a further physical understanding of the interplay of anionic quasi-atom and interstitial point defects in electrides, we show the results of high defects density using an 1×1×1 cubic FCC cell at here (defect density increases from 1/324 to 1/12) although this case is non-existent in our daily life.

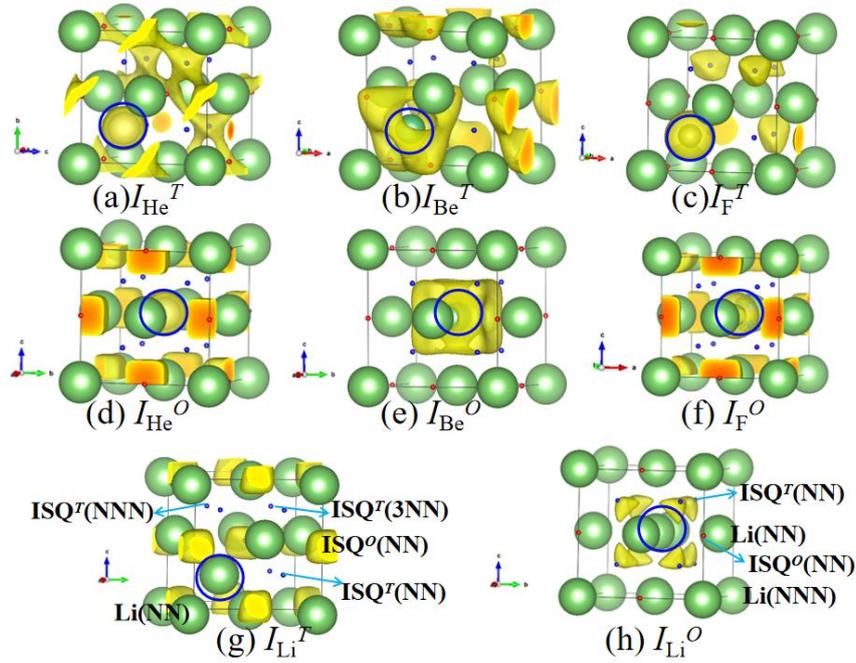

**Figure S26.** Calculated ELF (isosurface=0.75) of defective FCC-Li with different interstitial impurity in a cubic 1×1×1 FCC cell at 10 GPa. The big green, small red and small blue spheres represent Li atom, *O* and *T* sites, respectively. The point defects are marked with a blue circle.

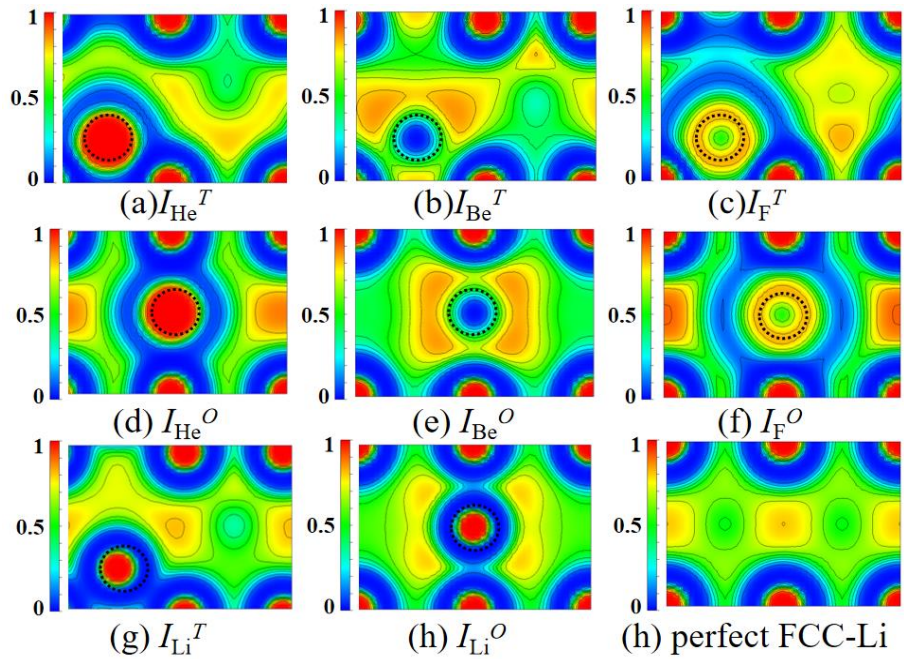

**Figure S27.** Calculated ELF in the (1 1 0) plane of defective FCC-Li as shown in **Fig. S26**, which are compared with perfect FCC-Li. The point defects are marked with black circles.

Figure S26 shows the ELF of these defective systems at 10 GPa, and the ELF of the (1 1 0) plane of these systems are shown in Fig. S27. We can see that ELF in these 1×1×1 cubic cells is similar with that in a 3×3×3 cubic supercell. However, there are obvious localization electrons at the $T$ sites. In addition, the ELF of $I_{\text{He}}^T$ shows more visible 3-D network characters. The Bader charge variations of matrix Li atoms, impurity atom, anionic ISQ$^O$ and ISQ$^T$ are shown in Figs. S28.

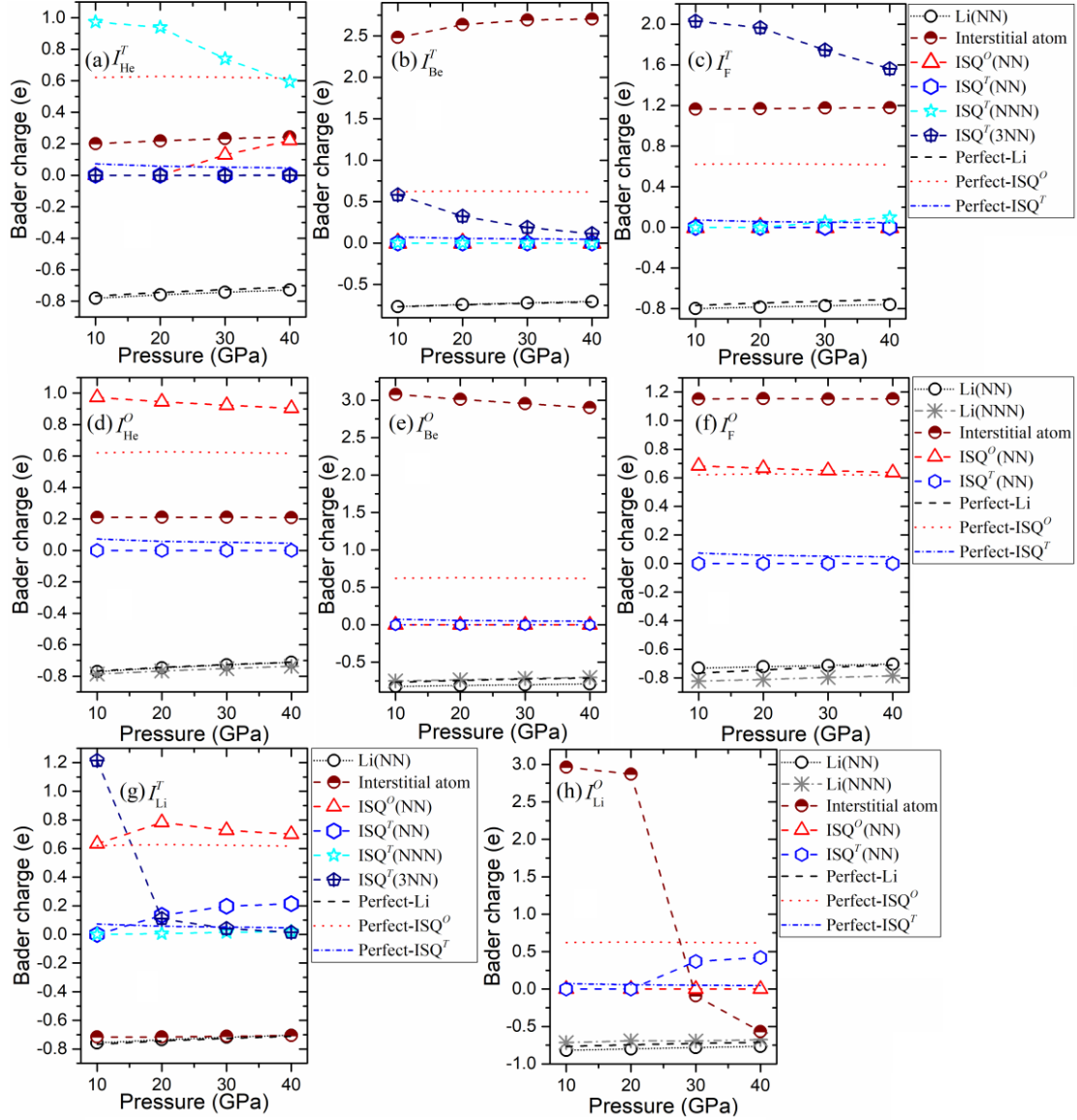

**Figure S28.** Variation of the Bader charge of the matrix Li atoms, impurity atom, anionic ISQ$^O$ and ISQ$^T$ of the 1×1×1 cubic FCC cell defective systems: (a) $I_{\text{He}}^T$, (b) $I_{\text{Be}}^T$, (c) $I_{\text{F}}^T$, (d) $I_{\text{He}}^O$, (e) $I_{\text{Be}}^O$, (f) $I_{\text{F}}^O$, (g) $I_{\text{Li}}^T$, (h) $I_{\text{Li}}^O$.

Compared with the results of a 3×3×3 cubic supercell, the increase of the defect density has no obvious influence on the valence state of matrix Li atoms, and it has the similar influence on the ISQs except for the self-interstitial Li defects. For example, the variations of the Bader charge for system $I_{He}^{T}$ also show a decrease of ISQ$^O$ (i.e. ISQ$^O$(NN)) and an increase of ISQ$^T$(NNN), and the Bader charge of ISQ$^O$ of the system $I_{Be}^{T}$ and $I_{F}^{T}$ are zero. However, the Bader charge of the $T$ interstitial sites (ISQ$^T$(NNN) in system $I_{He}^{T}$ and ISQ$^T$(NNN) in systems $I_{Be}^{T}$ and $I_{F}^{T}$ is large compared with the perfect FCC-Li, as well as compared with the defective FCC-Li with a low defect density. Obviously, this is a consequence of the strong interaction induced by the high impurity density.

For the cases of impurity atoms (He, Be, and F) occupying the $O$ sites, Bader charge variation of the ISQs in the 1×1×1 cubic cell is nearly the same as that in a 3×3×3 cubic supercell. That is to say, the density variation of the interstitial point defect occupying the $O$ sites cannot alter the electron localization characteristics nearby the impurity atoms.

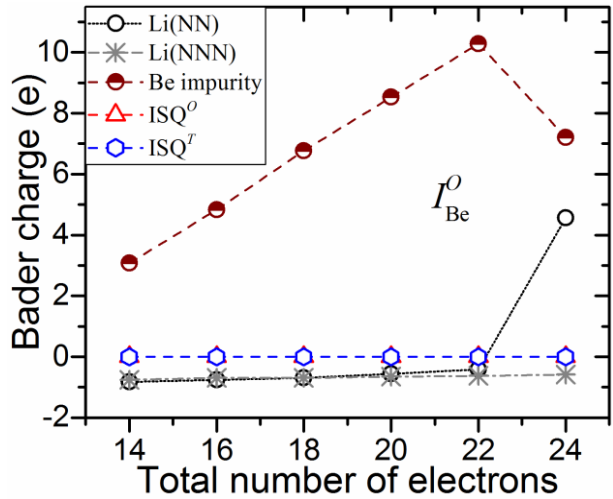

**Figure S29. Variation of the Bader charge of the matrix Li atoms, impurity Be atom, anionic ISQ$^O$ and ISQ$^T$ of $I_{Be}^{O}$ at 10 GPa with different total numbers of valence electrons modifying artificially.**

The charge state of impurity He and F are independent of the defect density. The approximate nominal valence state of He and F is -0 and -1, respectively. The approximate nominal valence state of Be is -3 in the 1×1×1 cubic cell, which is very different from that in a 3×3×3 cubic supercell. In these studied systems, the excess electrons localized in the interstitial sites only come from the matrix Li atoms. In other words, the total number of the excess electrons is a key quantity. In this regard, the decreases of the Bader charge of Be in a 1×1×1 cell results from the increase of the defect density. This is verified by our further Bader charge analysis. As shown in Fig. S29, the Bader charge of impurity Be in a 1×1×1 cell will increases when we add valence electrons in this system artificially. The fact that Be achieve the valence state of Be$^{8-}$ when we put $I_{Be}^{O}$ into a 4×4×4 FCC supercell also support this argument.

For system $I_{Li}^{T}$, the most obvious change in Bader charge is the sites of ISQ$^O$, ISQ$^T$(NN), and ISQ$^T$(3NN). This is similar to that in a 3×3×3 cubic supercell. But the change in the system $I_{Li}^{O}$ is very different if with a low defect density. In a 3×3×3 cubic supercell, the Bader charge of ISQ$^O$(NN) increases from 0 to 0.64$e$ at 20-30 GPa, the Bader charge of ISQ$^T$(NN) decreases sharply from 1.10 to 0.11$e$ at 20-30 GPa, the Bader charge of interstitial impurity Li maintains constant ~-0.7$e$ between 10-40 GPa. However, in an 1×1×1 cubic cell, the Bader charge of ISQ$^O$ (i.e. ISQ$^O$(NN)) maintains constant 0$e$ between 10-40 GPa, the Bader charge of ISQ$^T$ (i.e. ISQ$^T$(NN)) increases from 0 to 0.37$e$ at 20-30 GPa, the Bader charge of interstitial impurity Li decreases from 2.9$e$ (at 20 GPa) to -0.09$e$ (at 30 GPa), then to -0.56$e$ (at 40 GPa). This indicates that there are very strong interaction between Li atoms and ISQs, which has an obvious pressure dependence. This is supported by the variation of the ELF with pressure. As shown in Fig. S30, the electron localization become weaker in $T$ sites and become stronger in $O$ sites then the pressure increases. The volume enclosed by the ELF (isosurface=0.75) also decreases from 1.61 Å$^3$ (at 10 GPa) to 0.74 Å$^3$ (at 30 GPa).

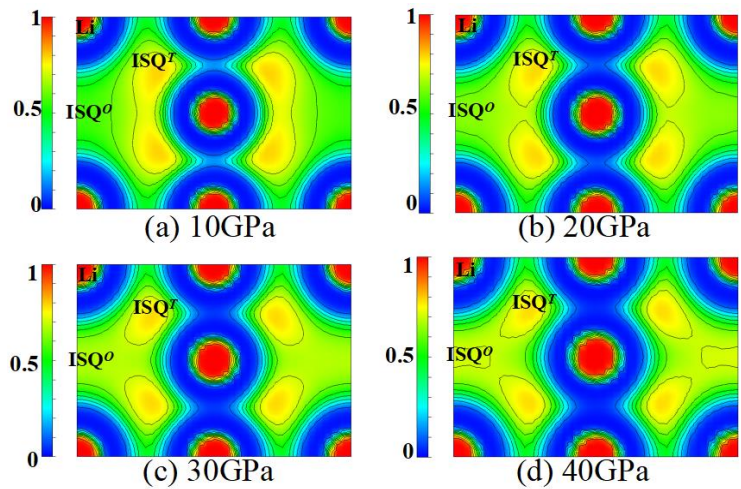

**Figure S30.** Calculated ELF in the (1 1 0) plane of system $I_{Li}^{O}$ at different pressures.

We know this case is non-existent in the real materials, but we think these results may be helpful for experimentalist to design new electrides and widen the applications of this class of materials.

## Supplementary references